\documentclass{aa}  

\usepackage{graphicx}
\usepackage{txfonts}

\newcommand{\tn}{\textnormal}

\begin{document} 

\title{Binary white dwarfs and decihertz gravitational wave observations: From the Hubble constant 
to supernova astrophysics.}


   \author{A. Maselli
          \inst{1}
          \and
          S. Marassi\inst{1}\and
          M. Branchesi\inst{2}}

   \institute{Dipartimento di Fisica, ``Sapienza'' Universit\`a di Roma \& Sezione INFN Roma1, 
   		Piazzale Aldo Moro 5, 00185, Roma, Italy\\
              \email{andrea.maselli@roma1.infn.it}, 
              \email{stefania.marassi@roma1.infn.it}
         \and
             Gran Sasso Science Institute (GSSI), I-67100 L’Aquila, Italy
		INFN, Laboratori Nazionali del Gran Sasso, I-67100 Assergi, Italy\\
             \email{marica.branchesi@uniurb.it}
             }


 
  \abstract
   {Coalescences of binary white dwarfs represent a copious source of information for gravitational wave 
   interferometers operating in the decihertz band. Moreover, according to the double degenerate scenario, 
   they have been suggested to be possible progenitors of supernovae (SNe) Type Ia events.}
   {In this paper we discuss the detectability of gravitational waves emitted by the inspiral of double white dwarfs. 
   We focus on the constraints that can be derived on the source's luminosity distance, and on other binary's parameters, 
   such as the angular momentum orientation.}
   {We explore the possibility of coincident detections of gravitational and electromagnetic signals; the 
   latter comes from the observation of the supernova counterpart. Confirmation of the double degenerate scenario 
   would allow one to use distances inferred in the gravitational wave channel to consistently calibrate SNe as standard 
   candles.}
   {We find that decihertz gravitational wave interferometers can measure the luminosity distance with relative accuracy better 
   than $1\%$ for binaries at 100 Mpc.  We show how multimessenger observations can put strong constraints on the 
   Hubble constant, which are tighter than current bounds at low redshift, and how they can potentially shed new light on the differences 
   with early-universe measurements.}
   {}


   \maketitle
%

\section{Introduction}

Relativistic compact sources, such as black holes, neutron stars, and white dwarfs, represent natural laboratories 
to probe fundamental laws of physics \citep{Barack:2018yly}. Gravitational wave (GW) observations of such objects 
have paved the way for a new understanding of the most extreme events of our Universe.
\citep{Abbott:2016nmj,Abbott:2016blz,Abbott:2017vtc,Abbott:2017oio,TheLIGOScientific:2017qsa,LIGOScientific:2018mvr}.\ These studies have aimed to address a large variety of open problems, which range from the status of matter at supranuclear density, 
to the nature of gravity in the strong field regime, and the dynamics of the expanding Universe \citep{Sathyaprakash:2019yqt}.

Among compact sources, binary white dwarfs (BWDs) feature a frequency 
content that is too low for the kilohertz sensitivity band of ground based detectors at the end of their orbital
evolution. Conversely, BWDs will be 
sources of GWs in the millihertz regime for LISA, either as an unresolved background or as individually resolvable 
events \citep{Nelemans:2001}.
Moreover, BWDs represent a golden target for GW decihertz interferometers \citep{Littenberg:2019grv}, 
operating between 0.1 and 1 Hz. These detectors can bridge the gap among existing 
and future ground-based facilities \citep{Akutsu:2018axf,Punturo:2010zz,Dwyer:2014fpa,Evans:2016mbw} and 
the millihertz range spanned by LISA \citep{Audley:2017drz,McWilliams:2019fng}. 

The GW decihertz interferometers represent unique laboratories to investigate the features of new astrophysical phenomena for the evolution of intermediate mass 
black holes \citep{Yagi:2012gb}, either in equal mass binaries or with a stellar companion, the stochastic background produced by 
cosmological sources \citep{Marassi:2009,Kowalska:2012}, or the nature of dark matter candidates \citep{Kawamura:2011zz}. 
Moreover, they have been proposed as a new tool to exploit multiband GW observations of stellar mass sources (either black holes or 
neutron stars).\ They are able to provide precise measurements of the source's localization \citep{Nair:2018bxj} and of the nuclear matter 
equation of state \citep{Isoyama:2018rjb}, and they allow for tests to be performed for gravity in the strong field regime \citep{Carson:2019rda,Gnocchi:2019jzp}.

Observations of BWDs by decihertz detectors would also shed new light on the evolutionary path of one of the most 
energetic events of our Universe, Type Ia supernovae \citep{Mandel:2017pzd,Sedda:2019uro}. Currently, two possible 
scenarios are believed to provide a major explanation of such phenomena \citep{1981NInfo..49....3T,Wang:2018} (see 
\citep{Wang:2012za,Maoz:2013hna} for a recent review). 
In the so called "single degenerate scenario", WDs accrete from a main sequence or a giant companion 
and eventually reach the mass-threshold to explode \citep{1973ApJ...186.1007W,Nielsen:2013kia}. 
According to the "double degenerate scenario", two WDs instead evolve through the emission of gravitational waves 
up to the merger phase, leading to the formation of single white dwarfs that are massive enough to ignite the burst 
\citep{1984ApJ...277..355W,1984ApJS...54..335I,2012ApJ...746...62R}. So far, electromagnetic 
surveys have not been able to distinguish between these two cases \citep{Rebassa:2019}. The GW signals 
in the decihertz regime can play a key role to probe nature and properties of the progenitors of 
supernovae (SNe) type Ia. In particular, coincident detections of SNe Ia and of GWs emitted by BWDs would provide 
the smoking gun needed to 
assess the consistency of the double degenerate scenario \citep{Mandel:2017pzd,Sedda:2019uro}. 

Type Ia SNe also play a crucial role as cosmic ladders. Electromagnetic observations of their light curves and spectra 
allow one to determine the luminosity distance and the redshift and, therefore, we are able to determine the parameters 
of the underlying cosmological model and the local value of the Hubble constant, in particular \citep{Hamuy:1996ss,Riess:1998cb,Kim:1998xz}. 
Current measurements of the Hubble constant can be classified into late-universe \citep{Riess:2019cxk,Wong:2019kwg} and 
early-universe \citep{Macaulay:2018fxi,Addison:2017fdm,Aghanim:2018eyx}  estimates, which lead in turn to high and 
low values of $H_0$, respectively. The tension between these two classes may be traced back into uncontrolled 
systematics, or as the emergence of new physics in one, or both, of the two 
regimes \citep{Riess:2019cxk,Aylor:2018drw}.

Gravitational waves have enabled us to potentially address this problem. Coalescing  
compact binaries represent  pure standard sirens: The GW emitted by these systems do not require any 
calibration, and allow one to uniquely determine the luminosity 
distance of the source \citep{Holz:2005df,Dalal:2006qt,1986Natur.323..310S,Krolak1987}. Coincident observations 
of such events and of their electromagnetic counterparts would allow one to disentangle the distance information with the 
binary's redshift, providing the value of the Hubble constant and of the other cosmological parameters 
\citep{PhysRevD.48.4738,Finn:1992xs}. GW170817 and the associated electromagnetic counterpart represent a 
genuine example of multimessenger astrophysics that has already led to a first estimate of $H_0$ 
\citep{TheLIGOScientific:2017qsa,Monitor:2017mdv,Abbott:2017xzu}. 
Although the majority of GW measurements are not accurate enough to enable a direct identification of the host 
galaxy, and thus of the redshift, future samples of joint electromagnetic and gravitational-wave detections can resolve 
the Hubble tension \citep{Nissanke:2009kt,Chen:2017rfc,Feeney:2018mkj}. 

In this paper we study the gravitational wave emission by coalescing binary white dwarfs. We analyze the accuracy 
on the source parameters that can be derived by decihertz GW signals and the possibility to use joint 
supernovae Ia observations (within the double degenerate scenario) to improve such constraints. 

\begin{figure}[th]
\centering
\includegraphics[width=8cm]{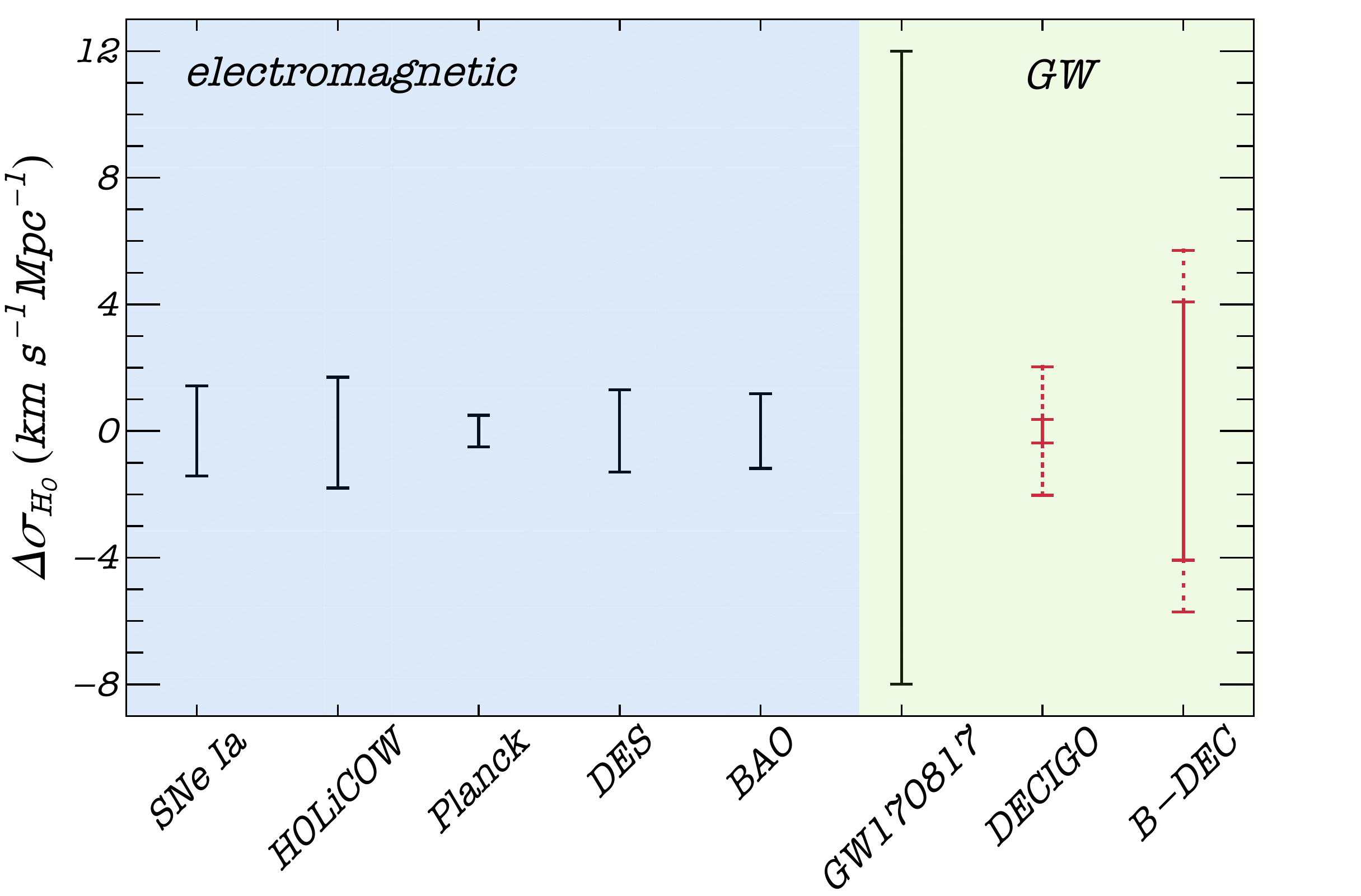}
\caption{1-$\sigma$ interval on the local value of the Hubble constant $H_0$ obtained by different astrophysical 
observations in the electromagnetic and gravitational wave bands. We consider constraints derived by SNe Ia events 
\citep{Riess:2019cxk}, the H0liCOW experiment \citep{Wong:2019kwg}, the Planck mission \citep{Aghanim:2018eyx}, 
the Dark Energy Survey \citep{Macaulay:2018fxi}, analysis of BAO \citep{Addison:2017fdm}, and the first measurement 
computed by coincident detection of the neutron star merger GW170817 and its EM counterpart \citep{Abbott:2017xzu}. 
The last two values correspond to specific bounds derived in this work for binary WDs observed in the decihertz band by 
DECIGO and B-DECIGO (see Table~\ref{tab:errors}-\ref{tab:errors2}). Dashed and solid lines correspond to errors 
taking the correction due to peculiar velocity into account.} 
\label{figs:bounds} 
\end{figure}

We show how, in a multimessenger scenario, BWDs are a complementary and independent tool 
to measure $H_0$ and are able to provide new insights on the fundamental physics of the SN explosion. 
Gravitational signals are assumed to be observed by the Japanese decihertz interferometers DECIGO and B-DECIGO 
\citep{Sato:2017dkf,Isoyama:2018rjb,Yagi:2011wg}. We focus on the constraints that these detectors will be able 
to derive on the luminosity distance and the Hubble constant, by exploiting the redshift inferred by the spectrum of SN 
or its host galaxy. Figure~\ref{figs:bounds} shows a summary of our results for $H_0$ compared against some of 
the existing constraints. The bounds derived in this paper are tighter than those currently 
available from electromagnetic surveys at small $z,$ and they are competitive with early Universe estimates that come, for example, 
from the cosmic microwave background \citep{Aghanim:2018eyx}.
The precise measurements of the luminosity distance can also be used to consistently calibrate the supernova light curves,
as recently proposed by \cite{Gupta:2019okl} in the context of double neutron star mergers.
We also show how when using GW signals emitted by BWDs, decihertz interferometers are able to accurately measure 
the inclination of the binary's orbital plane. The latter, together with the two component masses measured by GW emission, 
would provide a unique description of the binary's morphology, characterizing its full evolution, from the inspiral 
to the supernovae explosion. If not explicitly stated, throughout the paper we use geometrized units, in which $G = c = 1$.

\section{Waveform's model}

We consider the orbital evolution of binary white dwarfs up to the merger phase. 
In the frequency domain, the GW signal emitted during the inspiral can be 
described by the following post-Newtonian (PN) expanded GW waveform: 
\begin{equation}
\tilde{h}(f)=C_{\hat{\Omega}}\frac{\sqrt{3}}{2}\sqrt{\frac{5}{24}}\frac{G^{5/6}{\cal M}^{5/6}}{c^{3/2}\pi^{2/3}d}f^{-7/6}e^{i(\psi_\tn{PN}
-\psi_\tn{D}-\psi_\tn{pol})}\ ,\label{signal}
\end{equation}
where $d$ is the luminosity distance of the binary, and $(t_c,\phi_c)$ is the time and 
phase at the coalescence. The PN phase $\psi_\tn{PN}$ depends\footnote{For the purposes of this work, 
we consider a 2 PN phase, which includes both the chirp mass and the symmetric mass ratio. We neglect 
spins and, in general, finite size effects of the WDs \citep{Damour:2012yf,Vines:2011ud}.} on the chirp 
mass ${\cal M}=(m_1 m_2)^{3/5}/(m_1+m_2)^{1/5}$ and on the symmetric mass ratio 
$\nu=(m_1 m_2)/(m_1+m_2)^2$ \citep{Khan:2015jqa}, $\psi_\tn{D}$ describes the 
doppler shift of the phase wave front between the interferometer and the reference frame fixed with the Earth (or 
Sun) barycenter, and $\psi_\tn{pol}$ is the polarization phase \citep{Cutler:1997ta}.  The factor $C_{\hat{\Omega}}$ 
encodes the information on the source localization with respect to the detector's reference frame:
\begin{equation}
C_{\hat{\Omega}}=\sqrt{F_\times^2(\hat{L}\cdot \hat{N})^2 +F_+^2[1+(\hat{L}\cdot \hat{N})^2]^2/4}\ ,
\end{equation}
where $\hat{L}\cdot \hat{N}$ identifies the angle between the line of sight $\hat{N}$ and the binary's angular 
momentum $\hat{L}$, the latter being specified in the orbital plane by the polar and azimuthal angles 
$(\bar{\theta}_\tn{L},\bar{\phi}_\tn{L})$. The interferometer's pattern functions $F_{\times,+}(\theta_\tn{S},\phi_\tn{S},\psi_\tn{S})$ are defined as:
\begin{align}
F_+=\frac{1+\cos^2\theta_\tn{S}}{2}&\cos2\phi_\tn{S}\cos2\psi_\tn{S}-\cos\theta_\tn{S}\sin2\phi_\tn{S}\sin2\psi_\tn{S}\ ,\nonumber\\
F_\times=\frac{1+\cos^2\theta_\tn{S}}{2}&\cos2\phi_\tn{S}\sin\psi_\tn{S}+\cos\theta_\tn{S}\sin2\phi_\tn{S}\cos2\psi_\tn{S}\ .
\end{align}
Here $(\theta_\tn{S},\phi_\tn{S})$ describe the location of the BWD in the sky and $\psi_\tn{S}$ is the polarization 
angle \citep{Apostolatos:1994mx}, all of which are defined in the detector reference frame.
The strong correlation between $d$ and the inclination of the binary makes it difficult to extract the luminosity distance of the source, in general. For ground based detectors, the lack of extra information 
coming from electromagnetic counterparts prevent precise measurements of $d$, even in case of multiple 
GW interferometers \citep{Usman:2018imj}.

The three angles defined above, $(\theta_\tn{S}, \phi_\tn{S}, \psi_\tn{S})$, actually depend on time, since the detector 
moves on the orbit around the Sun. When knowing the configuration of the interferometer with respect to the orbital plane, 
it is straightforward to express these quantities in terms of the BWD angular velocity and of constant angles 
$(\bar{\theta}_\tn{S},\bar{\phi}_\tn{S})$, which are defined in a fixed reference frame centered with the Sun \citep{Cutler:1997ta,Yagi:2011wg,Nair:2018bxj}. 
We refer the reader to Appendix \ref{sec:detectorsangle} for the full expression of time-dependent variables 
as a function of constant quantities. The template in eq.~\eqref{signal} therefore depends on nine parameters, 
namely $\vec{\lambda}=(t_c,\phi_c,\ln{\cal M},\ln\nu,d,\bar{\theta}_\tn{S},\bar{\phi}_\tn{S},\bar{\theta}_\tn{L},\bar{\phi}_\tn{L})$.

\subsection{Signal-to-noise ratio and errors}\label{sec:snr}

Given the gravitational waveform \eqref{signal}, the signal-to-noise ratio (S/N) for 
a given detector is given by the noise weighted inner product of $\tilde{h}(f)$, that is:
\begin{equation}
\rho^2=4\ \tn{Re} \int_{f_\tn{min}}^{f_{\tn{max}}}\frac{df}{S_n(f)}\vert \tilde{h} (f)\vert^2\ ,
\end{equation}
with $S_n(f)$ being the detector's noise spectral density \citep{Cutler:1997ta}, 
which is explicitly given for DECIGO and B-DECIGO in Appendix~\ref{sec:detectorsangle}. 
For the systems considered in this work, we assume that $f_\tn{max}$ corresponds 
to the contact frequency of the system when the orbital separation is equal to 
$r_\tn{WD}(m_1)+r_\tn{WD}(m_2)$, assuming that the radius of the WD is given by 
the semi-analytic relation \citep{Schneider:2000sg,1996A&A...309..179P,1972ApJ...175..417N}: 
\begin{equation}
\frac{r_\tn{WD}(m_i)}{R_\odot}=0.012\left[\left(\frac{m_i}{1.44M_\odot}\right)^{-\frac{2}{3}}-\left(\frac{m_i}{1.44M_\odot}\right)^{\frac{2}{3}}\right]^{1/2}\ ,\  {\small i=1,2}\ , 
\end{equation}
while $f_\tn{min}$ corresponds to the frequency $T=5$ years before the merger. 
Specifically, the time evolution of the GW radiation in terms of the frequency 
is given at the two PN order by:
\begin{align}
t(f)=t_c-\frac{5}{256}{\cal M}(\pi{\cal M} f)^{-8/3}\left[1+\frac{4}{3}\left(\frac{743}{336}+\frac{11}{4}\nu\right)x
-\frac{32\pi}{5}x^{3/2}\right.\nonumber\\
\left.+2\left(\frac{3058673}{1016064}+\frac{5429}{1008}\nu+\frac{617}{144}\nu^2\right)x^2\right]\ ,
\end{align}
where $x=(\pi M f)^{2/3}$ \citep{Berti:2004bd}. For a given set of masses and 
observing time $T$, the minimum frequency can be numerically found as the solution 
of $t(f_\tn{max})-t(f_\tn{min})=T$.

For high S/Ns, the statistical errors on the source parameters can be computed through the 
Fisher information matrix \citep{Vallisneri:2008cx}, defined as:
\begin{equation}
\Gamma_{ij}=4\ \tn{Re} \int_{f_\tn{min}}^{f_{\tn{max}}}\frac{df}{S_n(f)}\frac{\partial \tilde{h}^\star (f)}{\partial\lambda^i}\frac{\partial \tilde{h}(f)}{\partial \lambda^j} \ ,
\end{equation}
where $\tilde{h}^\star (f)$ is the complex conjugate of the waveform, 
$\lambda^i$ is the $i$-th  term of the parameter's vector, and the Fisher is computed at 
the  true values of $\vec{\lambda}$. In this limit, the probability distribution of 
$\vec{\lambda}$, for a given detector's output $s$ is proportional to $\Gamma_{ij}$, that is, 
$p(\vec{\lambda\vert s})\propto p_0(\vec{\lambda})\exp \left[-\frac{1}{2}\delta \lambda^i\ \Gamma_{ij}\ 
\delta \lambda^i\right]$, where $\delta \lambda^i$ is the measurement shift with respect 
to the real values and $p_0(\vec{\lambda})$ is our prior information on $\vec{\lambda}$ 
\citep{Poisson:1995jb}. Diagonal and off-diagonal elements of the inverse of the Fisher 
matrix correspond to the root mean square and the correlation's coefficients of the source 
parameters:
\begin{equation}
\sigma_{\lambda_i}=\sqrt{(\Gamma^{-1})_{ii}}\quad , \quad
C_{\lambda_i,\lambda_j}=\frac{(\Gamma^{-1})_{ii}}{\sqrt{\sigma_{\lambda_i} \sigma_{ \lambda_j}}}\ .
\end{equation}

Following \citep{Yagi:2011wg}, we assume that each of the triangle units of (B-)DECIGO can be effectively 
considered as a system of two L-shaped interferometers, with the second being rotated $45^\circ$ with 
respect to the first one. In this configuration the pattern function of the second detector is given by 
$F^{(2)}_{\times,+}=F^{(1)}_{\times,+}(\theta_\tn{S},\phi_\tn{S}-\pi/4,\psi_\tn{S})$, and we can introduce 
a total S/N of $\rho=\sqrt{\rho^2_{(1)}+\rho^2_{(2)}}$. In the same way, the errors on the source parameters 
are obtained by inverting the sum of the Fisher matrices, that is, $\sigma^2_{\lambda_i}=(\Gamma^{(1)}+\Gamma^{(2)})_{ii}$.

The last ingredient of our analysis is given by the prior information on $\vec{\lambda}$. In this work we 
consider two possible scenarios: (i) BWDs that are only observed in the GW band, (ii) coincidence detections 
of WD mergers in both the electromagnetic and the gravitational spectrum. In this second scenario, we 
assume that the binary evolves according to the double degenerate scenario, igniting a SN Ia explosion. 
This assumption allows one to constrain the source's redshift $z$ and its sky position through the 
two angles $(\bar{\theta}_\tn{S},\bar{\phi}_\tn{S})$. We assume that the statistical errors on the polar 
and azimuthal angles are such that we can effectively reduce the dimensionality of the Fisher matrix 
and therefore of the parameter's vector to $\vec{\lambda}=(t_c,\phi_c,\ln{\cal M},\ln\nu,d,\bar{\theta}_\tn{L},\bar{\phi}_\tn{L})$. 

It is important to note that, unlike short gamma ray bursts that are associated with neutron star mergers, supernova explosions are not 
expected to be beamed. Consequently, they do not provide any information on the inclination angle of the binary, that is, the values of $\bar{\theta}_\tn{L}$ and of $\bar{\phi}_\tn{L}$ \citep{Monitor:2017mdv}.

In a multimessenger scenario, given the error of the luminosity distance and the knowledge of $z$, 
it is straightforward to propagate the uncertainty of $H_0$. For small redshifts $z<0.05$ (200 Mpc) for those considered 
in this paper, the local value of the Hubble constant can be determined by the Hubble-Lemaitre law, $H_0\simeq z/d$, 
neglecting errors\footnote{We assume that an uncertainty as to the redshift, measured by the supernova electromagnetic 
observations, is subleading compared to the other quantities in the analysis.} on $z$ this yields 
$\sigma_{H_0}\simeq z \sigma_{d_{\tn{L}}}/d^2$.
An additional source of uncertainty as to the $H_0$ measurement is given by the peculiar velocities; when the source 
is relatively close to the observer, the random relative motions of the galaxies due to gravitational interaction with nearby 
galaxies and overdensities are not negligible and the measured recessional velocity needs to be corrected in order to obtain the 
Hubble flow velocity. The unknown peculiar velocity is included in our $H_0$ uncertainties by adding in quadrature 
an additional statistical uncertainty of $1$-$\sigma$=200 km/s, which is a typical uncertainty for the peculiar velocity correction 
\citep{Carrick:2015xza}.

\section{Results}\label{sec:results}

\begin{figure}[th]
\centering
\includegraphics[width=9cm]{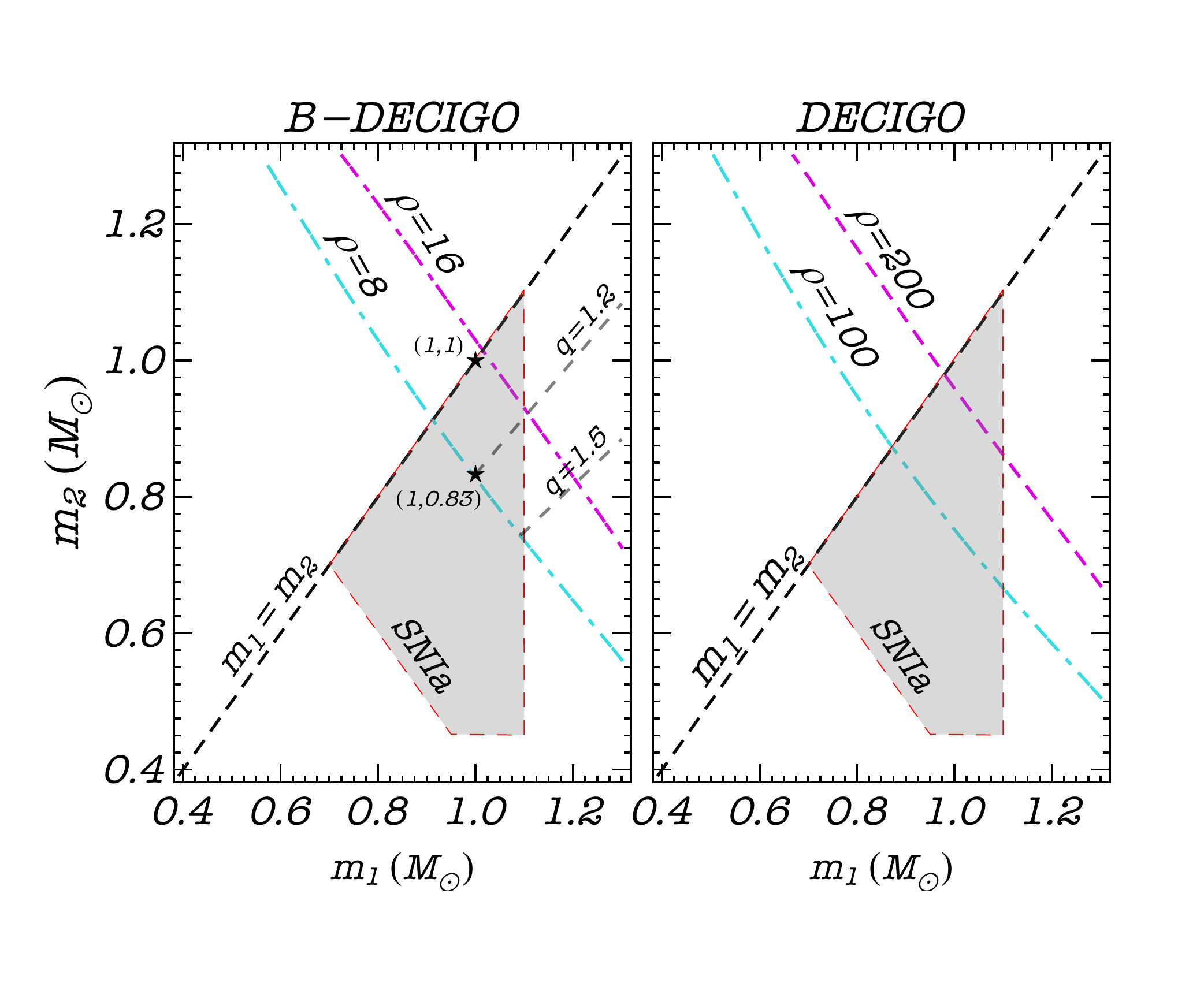}
\caption{Contour lines of fixed S/N $\rho$ as a function 
of the component masses $m_{1,2}$ for BWD observed by B-DECIGO (left panel) 
at $d=50$ Mpc and DECIGO (right panel) at $d=100$Mpc. The S/N was computed by
averaging over the source orientations. The shaded gray region identifies the 
region where the coalescence of the two stars may lead to supernovae Ia events 
\citep{Postnov:2014tza}. We also show lines of constant mass ratio $q=m_1/m_2$.} 
\label{fig:snr} 
\end{figure}

As a first step, we have investigated the observational window spanned by the 
interferometers for BWD systems. The two panels of Fig.~\ref{fig:snr} show the 
contour lines of fixed S/Ns as a function of the stellar masses for prototype 
binaries with $m_{1,2}\in [0.4,1.3]M_\odot$. Left and right plots correspond to 
systems observed by B-DECIGO and DECIGO at luminosity distances of $d=50$ Mpc 
and $d=100$ Mpc, respectively. We also averaged the GW signals over the sky 
localization and the polarization angles, knowing that 
$\langle C^2_{\hat{\Omega}}\rangle=4/25$. 
The shaded gray delimits the region where the two WD could lead to a SN type 
Ia event. For both the interferometers, a large portion of the parameter space 
does exist in which the binaries will be observed with a S/N above a threshold that, 
hereafter, we set to eight. For example, B-DECIGO will be able to 
detect gravitational signals emitted by the coalescence of a $1M_\odot$-$1M_\odot$ 
BWD up to $\sim 50$ Mpc. Fixing the primary mass $m_1$ and following the line of 
constant $\rho=8$, we see that binary configurations with a mass ratio of $m_1/m_2\leq 1.5$ 
are luminous enough to be resolved by B-DECIGO. At 100 Mpc, DECIGO will observe all of the 
systems  within the shaded area with a S/N above the threshold. This will be crucial 
to explore the whole mass spectrum of the BWD coalescence, in detail, and its 
connection with SN events.
We note that for the computation of the S/N, the distance acts as a 
scaling factor, namely $\rho\sim1/d$, and therefore our results can immediately be 
shifted to any value of $d$.  A more detailed study, as a function of the source's 
localization, is described in Appendix~\ref{sec:appsnr}.

The observations of SNe in the local Universe make it possible to derive volumetric 
rates for the different SN types as function of redshift. For the distances sampled by 
the present work, the volumetric rate of type Ia SNe is 
$0.25\pm 0.05 \times 10^{-4}\tn{Mpc}^{-3}\tn{yr}^{-1}$ \citep{Cappellaro:2015qia,Li:2010kd}, 
which corresponds to $\mathcal{O}(10)$ SNe Type Ia events per year within a volume of 50 
Mpc radius and $\mathcal{O}(10^2)$ for a volume of 100 Mpc radius.

\subsection{Accuracy of the source parameters}

Since the parameter's space to sample is rather large, we focus on the following: (i) equal 
mass binaries with $m_1+m_2=2M_\odot$; and (ii) specific values of the source distance 
and of the angular momentum's direction. For DECIGO (B-DECIGO), binaries are 
located $100$ Mpc  (50 Mpc) from the detector. 
Although the mass distribution of BWD is currently not known, we expect that 
our result will not be dramatically affected by small variations of the mass ratio 
$q=m_2/m_1$. \footnote{As an example, for binary configurations with $q\gtrsim1$, 
which is still compatible with the BWD-supernovae Ia scenario, the relative error of the luminosity 
distance $d$ would increase with the mass ratio, scaling at the leading order as 
$\sim\sqrt{q}$. We note, however, that for values of $q\gtrsim 1.2,$ the merger may not be 
able to ignite the SN explosion \citep{2011A&A...528A.117P}.}

The Fisher approach allows one to derive bounds on all the source parameters; however, 
hereafter we only discuss uncertainties about the luminosity distance and on the 
angles that identify $\hat{L}$. In particular, in order to investigate the accuracy on $\bar{\theta}_\tn{L}$ 
and $\bar{\phi}_\tn{L}$ we introduce the error box on the solid angle spanned by unit 
vector $\hat{L}$ as \citep{Cutler:1997ta}:
\begin{equation}
\Delta \Omega_\tn{L}=2\pi \vert\sin\bar{\theta}_\tn{L}\vert\sqrt{\sigma^2_{\bar{\theta}_\tn{L}}\sigma^2_{\bar{\phi}_\tn{L}}-
\Sigma^2_{\bar{\theta}_\tn{L}\bar{\phi}_\tn{L}}}\ .\label{errorO}
\end{equation}
We first consider the case without any prior on the localization. In this framework we 
assume that no electromagnetic counterpart has been observed in coincidence with the 
gravitational wave event and, according to Sec.~\ref{sec:snr}, we work with a full 
$9\times 9$ Fisher matrix. This scenario is expected to be more frequent\footnote{Possible 
absorption of the electromagnetic emission or large gravitational-wave localization may 
reduce the actual number of coincident detections \citep{Rebassa:2019}.}. 

The relative errors on the luminosity distance and on $\Delta \Omega_\tn{L}$ are shown 
in Table~\ref{tab:errors} for certain BWD's configurations, which feature S/Ns in the range of 
$\rho^\tn{DEC}\in(131, 290)$ and $\rho^\tn{B-DEC}\in(8, 19)$. Independent of the 
particular combination of $(\bar{\theta}_\tn{S},\bar{\phi}_\tn{S})$ and $(\bar{\theta}_\tn{L},\bar{\phi}_\tn{L}),$ 
the values of $\sigma_d/d$ obtained for DECIGO have a nearly flat distribution that clusters 
around $1\%$. For the smaller interferometer, the uncertainties on $d$ have roughly the same 
spread. At $50$ Mpc, we expect B-DECIGO to measure the luminosity distance of BWDs with 
10\% accuracy. Sources closer to the detector would improve these estimates.  As an example, 
a system at $d=10$ Mpc with the same sky location of the second binary in Table~\ref{tab:errors} 
leads to $\sigma^\tn{B-DEC}_d/d\simeq1\%$. It is also important to note that $d$ is almost uncorrelated with the 
other parameters, hence its error is proportional to the inverse of the S/N, namely 
$\sigma_d/d\sim1/\rho$, and our results can immediately be rescaled to any value of the 
luminosity distance. 

We can now focus on the accuracy on the orientation of the binary's angular momentum. 
The last two columns of Table~\ref{tab:errors2} show the projected constraints on 
$\Delta\Omega_\tn{L}$ defined in eq.~\eqref{errorO}. A satellite similar to DECIGO would estimate 
the direction of $\hat{L}$ with exquisite precision: For all of the models analyzed, we find a maximum 
error on $\Delta\Omega_\tn{L}$ smaller than one degree squared, even when considering sources 
at 100 Mpc from the detector. For B-DECIGO, the accuracy decreases of more than two orders of 
magnitude. At $d=50$ Mpc, in the best case scenario, $\Omega_\tn{L}$ can be determined with 
roughly $67$ deg$^2$ of accuracy. 

\begin{figure}[th]
\centering
\includegraphics[width=8cm]{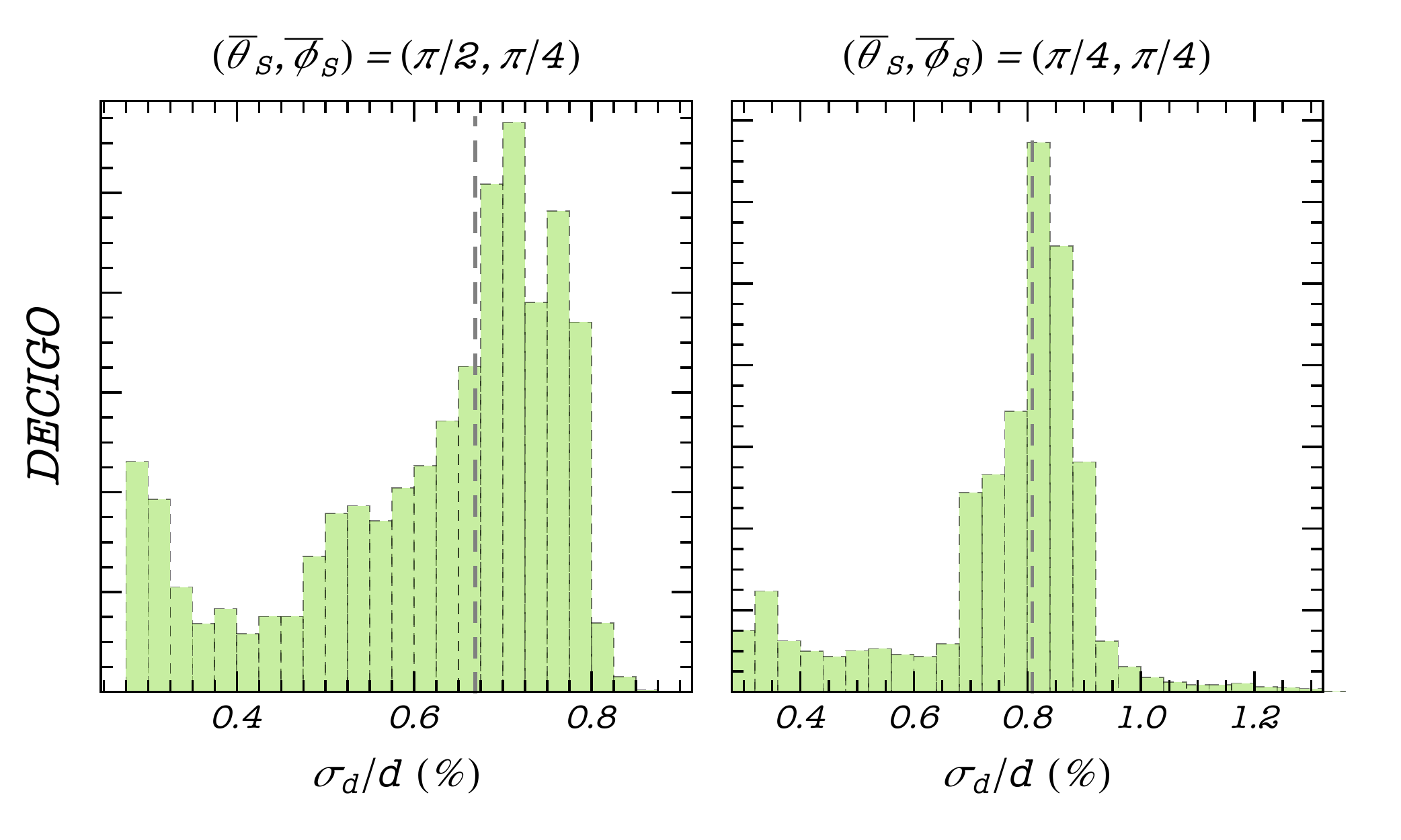}\\
\includegraphics[width=8cm]{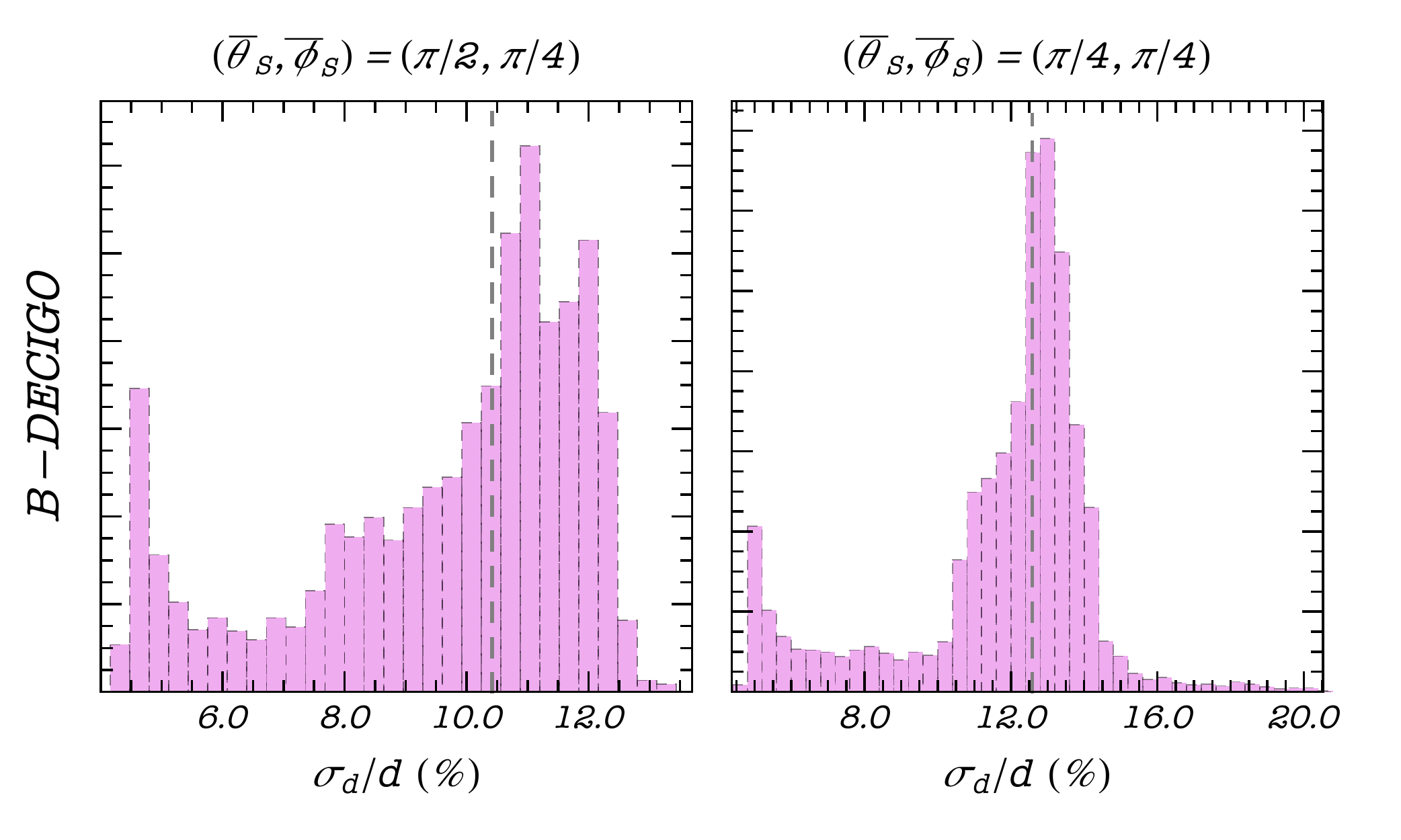}
\caption{1-$\sigma$ distributions of errors for the luminosity distance inferred 
for $n=10^4$ sources with orbital angular momentum randomly distributed on the sphere 
and a fixed sky localization with respect to the interferometers configurations 
$(\bar{\theta}_\tn{s},\bar{\phi}_\tn{s})=(\pi/2,\pi/4)$ left plots and $(\bar{\theta}_\tn{s},\bar{\phi}_\tn{s})=(\pi/4,\pi/4)$ 
right plots. Vertical dashed lines identify the median of the distribution. The binary systems are located at 
$d=100$Mpc and at $d=50$Mpc for DECIGO and B-DECIGO observations, respectively.
We consider equal mass WDs with $m_{1,2}=1M_\odot$.} 
\label{fig:errors1} 
\end{figure}

To further investigate the uncertainties on the luminosity distance and on the solid angle, we built an ensemble 
of $n=10^4$ BWDs in which $\cos\bar{\theta}_\tn{L}$ and $\bar{\phi}_\tn{L}$ are randomly drawn from uniform 
distributions within $[-1,1]$ and $[0,2\pi]$, respectively. Masses and distances are assumed to be the same as before, while 
polar and azimuthal angles are fixed to specific values. For each system we computed the corresponding covariance 
matrix. Figures~\ref{fig:errors1}-\ref{fig:errors2} show the histograms of the error's distributions on the luminosity 
distance for such configurations. The median of the 1-$\sigma$ for DECIGO are $\sigma_d/d\simeq 0.67\%$ and 
$\sigma_d/d\simeq 0.81\%$ for $\bar{\theta}_\tn{S}=\pi/2$ and $\bar{\theta}_\tn{S}=\pi/4$, respectively. Here, we set 
$\bar{\phi}_\tn{S}=\pi/4$, but the results are nearly identical for $\bar{\phi}_\tn{S}=0$.
For B-DECIGO, such a median increases to $\sigma_{d}/d\simeq 10.4\%$ and $\sigma_{d}/d\simeq 12.6\%$ 
for the two cases considered, namely $(\bar{\theta}_\tn{S},\bar{\phi}_\tn{S})=(\pi/2,\pi/4)$ and 
$(\bar{\theta}_\tn{S},\bar{\phi}_\tn{S})=(\pi/4,\pi/4)$. These values are in agreement with the single-source 
analysis shown in Table~\ref{tab:errors}. Finally, the two panels of Fig.~\ref{fig:errors3} show the cumulative density 
function of $\Delta\Omega_\tn{L}$ for the same set of $n=10^4$ binaries. For DECIGO, 90\% of the 
population yields $\Delta\Omega_\tn{L}\lesssim 1$ deg$^2$ for $\bar{\theta}_\tn{S}=\pi/2$, with the results 
being very similar for the case with $\bar{\theta}_\tn{S}=\pi/4$. The analysis for B-DECIGO leads to larger uncertainties, 
and the almost totality of the configurations sampled yield  errors of $\Delta\Omega_\tn{L}\gtrsim 100$ deg$^2$. We note that the knowledge of the angular momentum's direction represents a crucial piece of information that can be 
inferred from the binary's orbital motion. Assuming that the masses are also determined by the GW analysis, the 
measurement of $\hat{L}$ would help to reconstruct the full morphology of the system. For the multimessenger 
detections discussed below, this implies characterizing the BWD's evolution from the inspiral to the supernovae event. 

\begin{figure}[th]
\centering
\includegraphics[width=8.5cm]{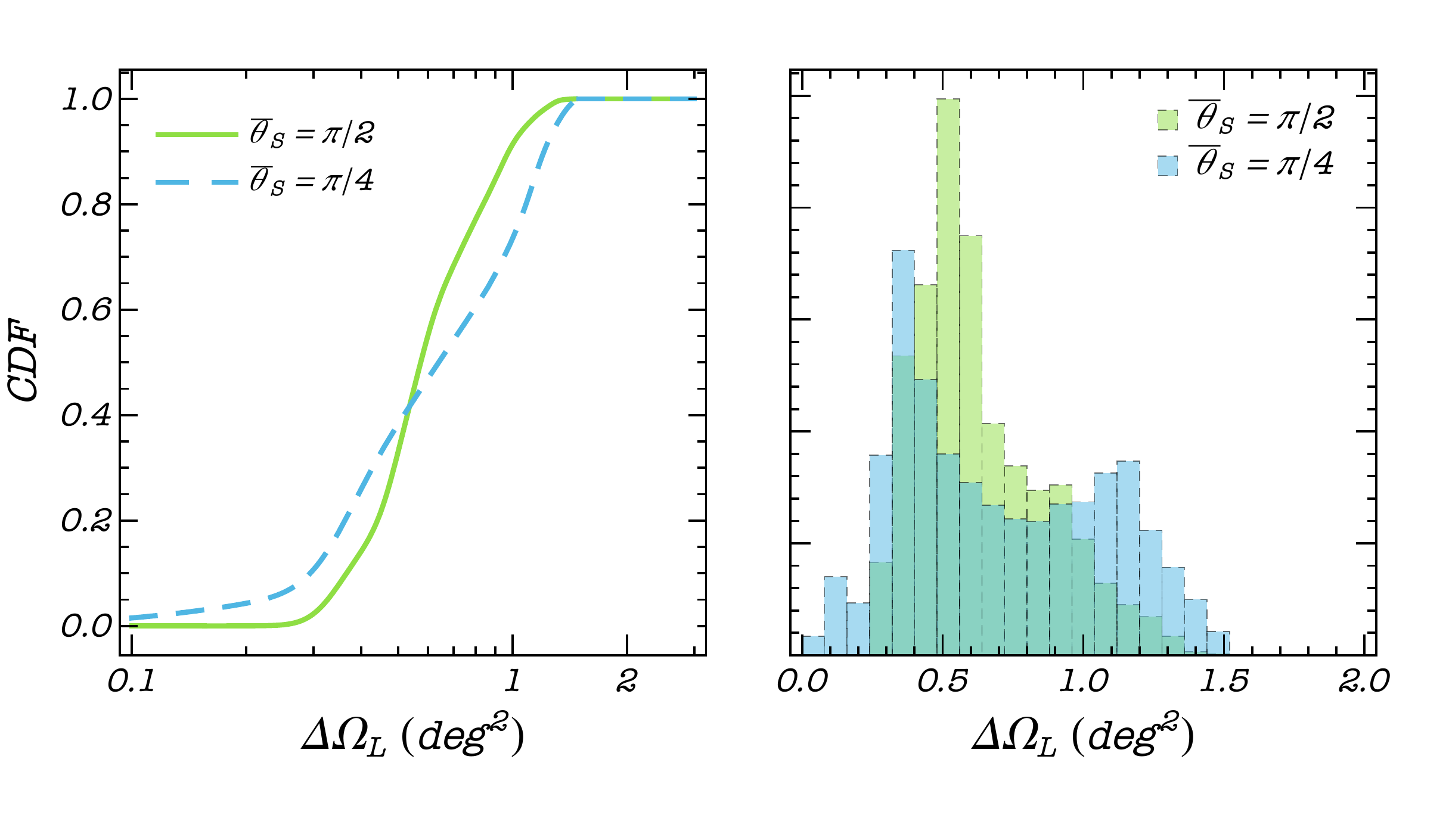}
\caption{Cumulative (left panel) and histogram (right panel) distribution 
of the 1-$\sigma$ errors on the solid angle $\Omega_\tn{L}$ 
derived from the ensemble of $n=10^4$ binary white dwarf with angular momentum 
randomly oriented and specific configurations of $(\bar{\theta}_\tn{S},\bar{\phi}_\tn{S})$. 
Sources are observed by DECIGO at $d=100$ Mpc.} 
\label{fig:errors3} 
\end{figure}

\begin{table*}[ht]
        \renewcommand*{\arraystretch}{1.1}
  \centering
     \begin{tabular}{cccccccc}
   \hline
   \hline
   $\cos\bar{\theta}_\tn{L}$ & $\bar{\phi}_\tn{L}$ & $\cos\bar{\theta}_\tn{S}$ & $\bar{\phi}_\tn{S}$ & $\sigma^\tn{DEC}_d/d\ \%$ 
   & $\sigma^\tn{B-DEC}_{d}/d\ \%$ & $\Delta\Omega_\tn{L}^\tn{DEC}$ & $\Delta\Omega_\tn{L}^\tn{B-DEC}$\\
 \hline
-0.2 & 4 & \phantom{-}0.3 & 5 & 0.787 (0.806) & 12.3 (12.6) & 0.899 (0.942) & 219 (229) \\
 \phantom{-}0.2 & 0 & \phantom{-}0.3 & 5 & 0.561 (0.562) & 8.75 (8.76) & 0.284 (0.298) & 69.2 (72.6) \\
 -0.2 & 4 & -0.3 & 1 & 0.391 (0.391) & 6.10 (6.10) & 0.379 (0.380) & 92.4 (92.6) \\
 \phantom{-}0.2 & 0 & -0.3 & 1 & 0.777 (0.796) & 12.1 (12.4) & 0.895 (0.938) & 218 (228) \\
 -0.2 & 4 & \phantom{-}0.3 & 3 & 0.563 (0.564) & 8.74 (8.76) & 0.289 (0.303) & 70.0 (73.5) \\
 \phantom{-}0.2 & 0 & \phantom{-}0.3 & 3 & 0.780 (0.812) & 12.1 (12.7) & 0.905 (0.958) & 219 (232) \\
 -0.2 & 4 & -0.3 & 6 & 0.542 (0.542) & 8.41 (8.41) & (0.278) (0.294) & 67.4 (71.2) \\
 \phantom{-}0.2 & 0 & -0.3 & 6 & 0.768 (0.779) & 11.9 (12.1) & 0.838 (0.872) & 202 (211) \\
 -0.8 & 4 & \phantom{-}0.5 & 5 & 0.774 (0.774) & 12.1 (12.1) & 0.35 (0.353) & 85.0 (85.9) \\
 \phantom{-}0.8 & 0 & \phantom{-}0.5 & 5 & 0.806 (0.816) & 12.6 (12.7) & 0.466 (0.486) & 114 (119) \\
 -0.8 & 4 & -0.5 & 1 & 0.741 (0.742) & 11.6 (11.6) & 0.408 (0.429) & 98.7 (104) \\
 \phantom{-}0.8 & 0 & -0.5 & 1 & 0.762 (0.762) & 11.9 (11.9) & 0.357 (0.360) & 86.9 (87.9) \\
 -0.8 & 4 & \phantom{-}0.5 & 3 & 0.797 (0.805) & 12.4 (12.5) & 0.409 (0.424) & 98.8 (102) \\
 \phantom{-}0.8 & 0 & \phantom{-}0.5 & 3 & 0.749 (0.749) & 11.6 (11.7) & 0.352 (0.358) & 84.8 (86.4) \\
 -0.8 & 4 & -0.5 & 6 & 0.783 (0.792) & 12.2 (12.3) & 0.49 (0.508) & 118 (123) \\
 \phantom{-}0.8 & 0 & -0.5 & 6 & 0.760 (0.760) & 11.8 (11.8) & 0.323 (0.325) & 77.9 (78.4) \\
    \hline
  \end{tabular}  
  \caption{1-$\sigma$ uncertainties of the luminosity distance $d$ (relative percentual) 
  and of the solid angle (deg$^2$) identified by the binary angular momentum for certain 
  combinations of $(\bar{\theta}_\tn{S},\bar{\phi}_\tn{S})$ and of the angles $(\bar{\theta}_\tn{L},\bar{\phi}_\tn{L})$. 
  White dwarf binaries have masses $m_{1,2}=1M_\odot$, and are placed at $d=100$ 
  Mpc and $d=50$ Mpc, from DECIGO and B-DECIGO, respectively. Values between 
  round brackets correspond to errors evaluated for the same configuration through the 
  multimessenger analysis, i.e. knowing the source localization given by $
  (\bar{\theta}_\tn{S},\bar{\phi}_\tn{S})$.}
\label{tab:errors} 
\end{table*}

\subsection{Multimessenger detections and the Hubble constant}

\begin{figure}[th]
\centering
\includegraphics[width=8.5cm]{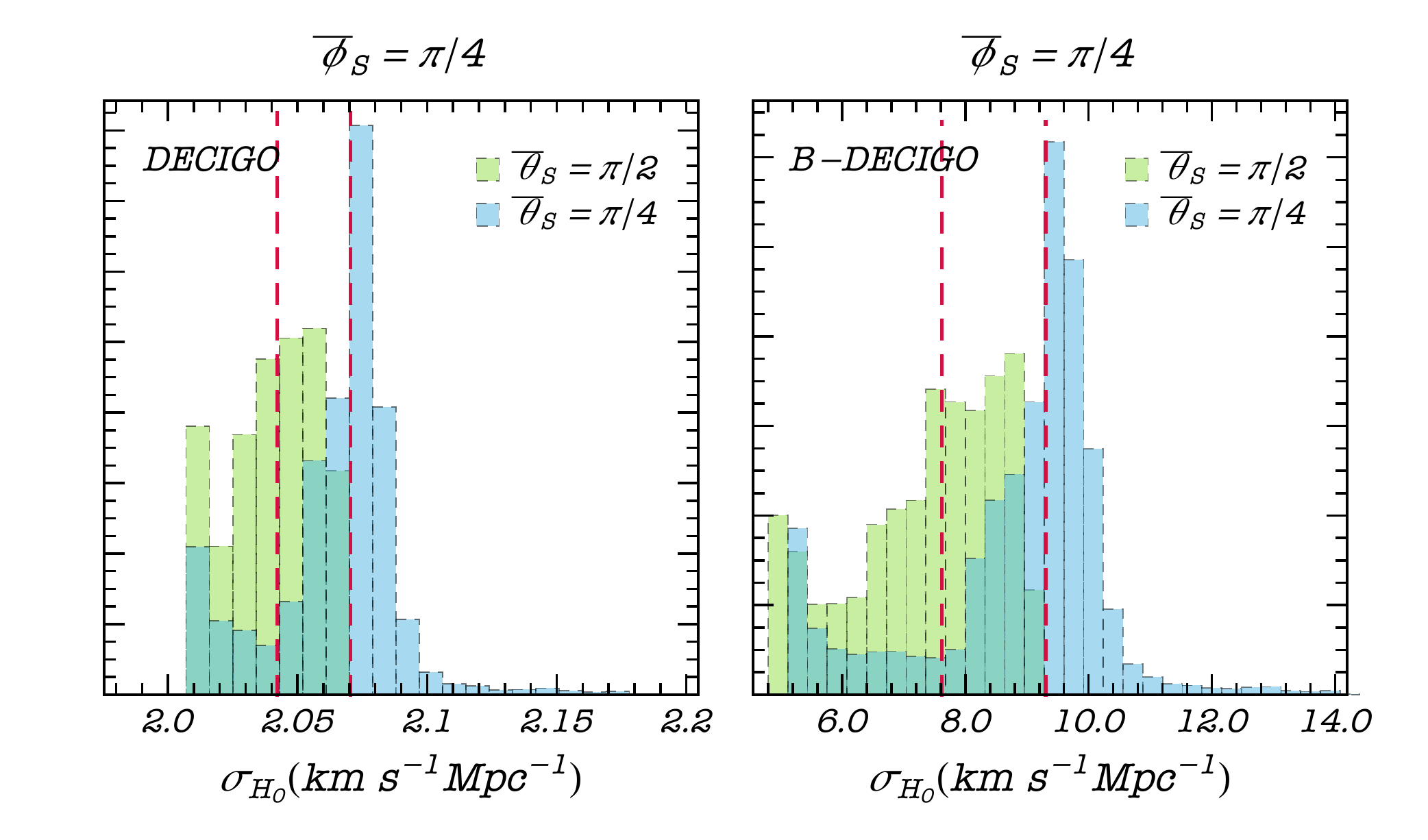}
\caption{Error distribution of 1-$\sigma$  for the Hubble constant computed for the set of $n=10^4$ 
binary white dwarfs also considered in Fig.~\ref{fig:errors1}.} 
\label{fig:errors2} 
\end{figure}

We now focus on the multimessenger scenario in which a SN Ia event is observed 
in coincidence with the GW signal. In this case, we assume that the binary's polar and 
azimuthal angles are known. In practice, we removed $\bar{\theta}_\tn{S}$ and $\bar{\phi}_\tn{S}$ 
from the Fisher matrix, reducing the number of parameters to constrain. The values of 
$\sigma_d/d$ and $\Delta\Omega_\tn{L}$ that were computed following this approach are shown 
between round brackets in Table~\ref{tab:errors}. 
For the angles chosen in Table~\ref{tab:errors}, the uncertainties of both the 
luminosity distance and the solid angle are close to those derived in the previous section. 
For both DECIGO and B-DECIGO, we obtain differences up to $4\%$ on 
$d$  and $6\%$ on $\Delta \Omega_\tn{L}$ between the 1-$\sigma$ derived with and 
without the supernova prior. We have repeated this analysis for the statistical ensemble 
of $10^4$ binaries shown in Figs.~\eqref{fig:errors1}-\eqref{fig:errors2}, and we find, however, 
that some particular combinations of the source orientation may lead to 
discrepancies on the order of $10\%$ on the luminosity distance and up to $100\%$ 
on the angular momentum direction. Multimessenger constraints therefore harbor the 
potential to significantly increase the accuracy of the parameter estimation. 

However, the observation of the electromagnetic counterpart is also crucial in disentangling the 
source's redshift. The latter can be used to translate constraints of the luminosity distance 
into bounds on the Hubble constant. We show such values in Table~\ref{tab:errors2}. The 
errors of $H_0$ that were computed for DECIGO are smaller than the local measurements inferred using 
supernovae observations \citep{Riess:2019cxk} or gravitational lensing time delays 
\citep{Wong:2019kwg}, and they are comparable with those obtained from the cosmic 
microwave background \citep{Aghanim:2018eyx} (see also Fig.~\ref{figs:bounds}). We also note that the exquisite precision of DECIGO makes it so $\sigma_{H_0}$ is dominated by the 
uncertainty as to the peculiar velocity which, at 100 Mpc, can be up to a factor $\sim2$ higher 
than the errors coming from the luminosity distance alone. As expected, uncertainties for 
B-DECIGO are, in general, larger. At 50 Mpc, the projected constraints are looser than current 
bounds in the electromagnetic band, although they are still smaller than the value derived 
for the first double neutron star GW event \citep{Abbott:2017xzu}. The last two columns 
of Table~\ref{tab:errors2} also show that for B-DECIGO, the uncertainty of the Hubble 
parameter is dominated by the statistical error coming from the GW parameter estimation.

The two panels of Fig.~\ref{fig:errors2} show the errors on $H_0$ that were computed for the random set of 
$10^4$ binaries introduced in the previous section. Sources with smaller values of $\bar{\theta}_\tn{S}$ 
lead to larger uncertainties for both the detectors. This picture does not change dramatically if 
we vary $\bar{\phi}_\tn{S}$. Overall, we find medians, which were corrected by the peculiar velocity,
for DECIGO of $\sigma_{H_0}\simeq 2.04$ km\ s$^{-1}$Mpc$^{-1}$ 
($\bar{\theta}_\tn{S}=\pi/2$) and $\sigma_{H_0}\simeq 2.07$ km\ s$^{-1}$Mpc$^{-1}$ ($\bar{\theta}_\tn{S}=\pi/4$). 
For B-DECIGO, such values increase to $\sigma_{H_0}\simeq 7.62$ km\ s$^{-1}$Mpc$^{-1}$ and 
$\sigma_{H_0}\simeq 9.30$ km\ s$^{-1}$Mpc$^{-1}$.
Therefore, in complete analogy with the coalescence of neutron stars and their gamma-ray-burst counterpart 
\citep{Abbott:2017xzu}, multimessenger observations of BWDs provide a powerful tool to determine the local value 
of the Hubble constant, which is independent and competitive with current constraints.

\begin{table}[ht]
        \renewcommand*{\arraystretch}{1.1}
 \addtolength{\tabcolsep}{-4pt}
  \centering
     \begin{tabular}{cccccccc}
   \hline
   \hline
   $\cos\bar{\theta}_\tn{L}$ & $\bar{\phi}_\tn{L}$ & $\cos\bar{\theta}_\tn{S}$ & $\bar{\phi}_\tn{S}$ & $\sigma_{H_0}^\tn{DEC}$ 
   & $\sigma_{H_0,vel}^\tn{DEC}$ & $\sigma_{H_0}^\tn{B-DEC}$ & $\sigma_{H_0,vel}^\tn{B-DEC}$ \\
 \hline
-0.2 & 4 & \phantom{-}0.3 & 5 & 0.522 & 2.07 & 8.21 & 9.13 \\
\phantom{-}0.2 & 0 & \phantom{-}0.3 & 5 & 0.372 & 2.03 & 5.85 & 7.09 \\
 -0.2 & 4 & -0.3 & 1 & 0.259 & 2.02 & 4.08 & 5.71 \\
 \phantom{-}0.2 & 0 & -0.3 & 1 & 0.515 & 2.07 & 8.08 & 9.02 \\
 -0.2 & 4 & \phantom{-}0.3 & 3 & 0.373 & 2.03 & 5.84 & 7.08 \\
 \phantom{-}0.2 & 0 & \phantom{-}0.3 & 3 & 0.517 & 2.07 & 8.11 & 9.05 \\
 -0.2 & 4 & -0.3 & 6 & 0.359 & 2.03 & 5.62 & 6.90 \\
 \phantom{-}0.2 & 0 & -0.3 & 6 & 0.509 & 2.06 & 7.98 & 8.93 \\
 -0.8 & 4 & \phantom{-}0.5 & 5 & 0.513 & 2.06 & 8.08 & 9.02 \\
 \phantom{-}0.8 & 0 & \phantom{-}0.5 & 5 & 0.534 & 2.07 & 8.42 & 9.32 \\
 -0.8 & 4 & -0.5 & 1 & 0.492 & 2.06 & 7.75 & 8.72 \\
 \phantom{-}0.8 & 0 & -0.5 & 1 & 0.505 & 2.06 & 7.94 & 8.89 \\
 -0.8 & 4 & \phantom{-}0.5 & 3 & 0.528 & 2.07 & 8.29 & 9.20 \\
 \phantom{-}0.8 & 0 & \phantom{-}0.5 & 3 & 0.496 & 2.06 & 7.78 & 8.75 \\
 -0.8 & 4 & -0.5 & 6 & 0.519 & 2.07 & 8.14 & 9.07 \\
 \phantom{-}0.8 & 0 & -0.5 & 6 & 0.504 & 2.06 & 7.90 & 8.85 \\
      \hline
  \hline
  \end{tabular}  
  \caption{Uncertainties of the Hubble constant $H_0$ (in  km s$^{-1}$Mpc$^{-1}$) evaluated 
  by coincident electromagnetic and gravitational wave detections of the same binary 
  configurations shown in Table~\ref{tab:errors}. For each configuration, we show the error 
  corrected with and without the velocity dispersion.}
\label{tab:errors2} 
\end{table}

In addition to the cosmological application, coincident detections of merging WDs may also be used to calibrate SN Ia 
luminosities, as was recently proposed in \citep{Gupta:2019okl} by comparing the gravitational and  the electromagnetic 
measurements of $d$. In this work, the authors consider GWs that were emitted by 
neutron star mergers, occurring in galaxies that host supernovae explosions. This strategy, however, may suffer from 
spurious systematics, as there is a lack of precise knowledge on the relative position between the two events. In our 
approach, the association between the WDs merger and the SN Ia, through the double degenerate scenario, 
would provide a unique and consistent estimate of the source's luminosity distances and of its flux.

\section{Conclusions}

Along with black holes and neutron stars, white dwarfs represent one of the flavors 
in which compact objects manifest in the Universe. Similar to single or binary sources, 
their evolution leads to a rich 
phenomenology, which is connected with a large variety of astrophysical phenomena 
\citep{Postnov:2014tza,2015ApJ...805L...6S}. Depending on the nature of the companion, WD 
coalescences could ignite the emission of different electromagnetic counterparts, as X-transients 
\citep{Sesana:2008zc,Bauer:2017enl}. Moreover, binary WDs have received a lot of attention as possible 
progenitors of supernovae Ia events, according to the double degenerate scenario. For these reasons, 
they represent ideal candidates to fully exploit multimessenger observations.   

In this paper, we investigate the detectability of coalescing white dwarf systems at the end 
of their inspiral phase via decihertz gravitational wave interferometers.
We computed the S/N for prototype binaries, assessing the accuracy of the source's 
parameters estimated by the Japanese detectors DECIGO and B-DECIGO.
We primarily focus on the 
constraints that can be placed on the luminosity distance. Indeed, since the orbital evolution of a compact 
binary is completely determined by general relativity, BWDs are clean standard sirens. We find 
that DECIGO can measure the source's distance with $1\%$  accuracy and with better accuracy for binaries at 100 
Mpc from the detector. B-DECIGO is able to perform the same quality-measurements for systems one 
order of magnitude closer, that is, for $d<20$ Mpc.\ Although, the interferometer will still constrain the luminosity 
distance with a relative accuracy of $10\%$ within an horizon of $50$ Mpc.

We explore the multimessenger scenario, in which GWs signals are observed 
in coincidence with SN Ia events, with the latter providing the source's polar and azimuthal angles. 
Overall, we find a mild improvement of the statistical errors on $d$. 
However, the joint analysis has a crucial impact on cosmology: Assuming that the electromagnetic channel 
 only yields the binary's redshift, the local value of the Hubble constant can now be determined. Our results 
suggest that DECIGO (B-DECIGO) can put a bound on $H_0,$ such that at a $68\%$ confidence level, 
$\sigma_{H_0}\lesssim2.1$ ($\sigma_{H_0}\lesssim5.7$) km s$^{-1}$Mpc$^{-1}$ at $d=100$ Mpc ($d=50$ Mpc). 
The strategy devised in this paper is complementary 
to measurements at low redshift that are available 
nowadays and, therefore, it offers an independent approach to alleviate or solve the tension on the Hubble constant.

While for binary neutron stars, the beamed electromagnetic emission can constrain the orbital plane and thus improve 
the distance and the Hubble constant measurements \citep{2019NatAs.tmp..385H,2019PhRvX...9c1028C}; for BWD, 
the $H_0$ measurement precision benefits from a higher rate of joint detections, due to the isotropic emission of SN 
type Ia with respect to the GRB and a brighter emission with respect to a kilonova, and from a higher astrophysical rate 
of BWD with respect to the BNS rate (the white-dwarf merger rate is about 1-2 orders of magnitude higher than for BNS) 
\citep{Badenes:2012ak,Abbott:2017ntl}.
Indeed, the projected constraints discussed so far can be further improved by stacking multiple 
GW observations. To show how $\sigma_{H_0}$ changes, we simulated one thousand sets 
of 100 GW detections each one. For each binary the primary mass and mass ratio are drawn randomly 
from uniform distributions within the intervals $m_1\in[0.8,1.1]M_\odot$ and $q=m_2/m_1\in[1,1.2]$. The source orientations 
and angular momenta were also randomly chosen on the sphere.  
The luminosity distance of such events is picked from a uniform distribution between 50 and 
200 Mpc. We stacked a different number of observations, which shows the 
corresponding errors for DECIGO in Fig.\ 6. By comparing such values with the uncertainties found 
in Fig.~\ref{fig:errors1} and Table~\ref{tab:errors2}, we note that exploiting the white-dwarf 
merger rate would reduce the error on the Hubble constant 
by almost one order of magnitude with 100 detections.
 
\begin{figure}[th]
\centering
\includegraphics[width=5.5cm]{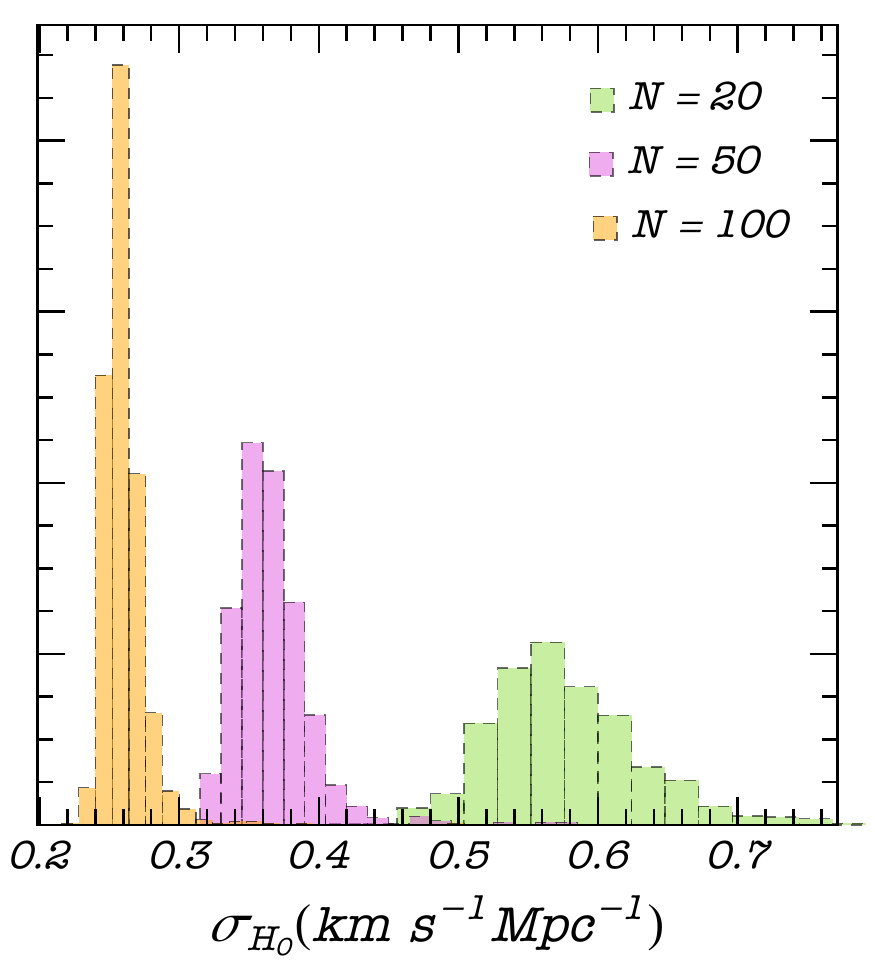}
\caption{Projected 1-$\sigma$ constraints on the value of the Hubble constant measured with 
DECIGO as a function of the number of GW observations.} 
\label{fig:errors3} 
\end{figure}

In addition to the cosmological implications, coincident detections of binary white dwarfs may be used to calibrate the 
supernova's light-curves, as recently suggested in \cite{Gupta:2019okl}. We remark that confirmation of the 
double degenerate scenario would provide a fully consistent  calibration of the SN flux, free of any systematics 
due to missing information between the electromagnetic and the gravitational event, which is in analogy with 
GW170817 \citep{Abbott:2017xzu}.

This statistical study also allowed us to determine the accuracy with which decihertz interferometers 
will be able to measure the inclination of the BWD's orbital plane. We find that DECIGO can constrain 
the solid angle identified by the orbital angular momentum with accuracies below 3 deg$^2$. This 
represents an important piece of information, which characterizes the binary's evolution prior 
the merger, and therefore it may have deep implications for the supernova explosion. 

The data analysis carried out in this paper does not take finite size effects due to rotation or 
tidal interactions into account \citep{Piro:2011qe,McNeill:2019rct,2012MNRAS.420.3126F}.  Such corrections affect the 
waveform at higher post-Newtonian order, and they are expected to be small compared to the dominant contribution 
\citep{Willems:2009xk,2012ApJ...745..137V}. However, we plan to improve upon the GW description and further explore 
the impact these modifications have in a forthcoming publication. 
As a final remark, due to the relevance of binary white dwarf within the stellar evolution, it would be interesting to study 
the source localization of such systems in the millihertz band spanned by LISA. A detailed study on this topic, which
also exploits a catalog of binaries sampled and adopts the approach described in 
\cite{Schneider:2017rdg}, \cite{Graziani:2017xtx} and \cite{Marassi:2019xjy}, is already 
ongoing \citep{Maselli:2019}.

\begin{acknowledgements}
A.M wants to thank Swetha Bhagwat and Luca Izzo for useful discussions and 
for having carefully read this manuscript. The Authors also thank the organizers 
of the PAX meeting, ``Physics and Astrophysics at the eXtreme", held in Cascina 
in 2019, where the idea of this work was discussed. 

A.M acknowledges support from the Amaldi Research Center funded by the MIUR 
program ``Dipartimento di Eccellenza'' (CUP: B81I18001170001). The authors would 
like to acknowledge networking support by the COST Action CA16104.
\end{acknowledgements}

\bibliographystyle{aa} 
\bibliography{biblio.bib}

\begin{appendix}

\section{Detector configuration}\label{sec:detectorsangle}
In this section we give a brief review of the parameters that characterize the orbital configuration 
of DECIGO and B-DECIGO. We refer the reader to \cite{Cutler:1997ta} for a detailed analysis of 
the source's localization by space interferometers with an instrumental design similar to the 
one considered in this paper.

DECIGO is planned to be composed of four clusters of three spacecrafts each. The latter are separated 
by 1000 km, forming a triangular configuration that moves on a Helio-centric orbit \citep{Sato:2017dkf}.
In this configuration, we identified two reference 
systems: one aligned with the detector, and one fixed with the barycenter. We attached the 
coordinates $(x,y,z)$ and $(\bar{x},\bar{y},\bar{z})$ to these two systems, respectively. The arms of the interferometer lie in the $x-y$ 
plane, while  the $z-$axis is inclined at an angle of $\gamma=\pi/3$ with respect to $\bar{z}$ and it precesses around 
the latter at a constant rate, such that:
\begin{equation}
z^i=\frac{\bar{z}^i}{2}-\frac{\sqrt{3}}{2}[\cos\bar{\phi(t)} \bar{x}^j+\sin\bar{\phi(t)} \bar{y}^j]\ ,\label{precession}
\end{equation}
where $(x^j,y^j,z^j)$ are the unit vectors along the $(z,y,z)$-axis (and similarly for the barred coordinate). 
The orbital motion of the satellite is specified by $\bar{\phi}(t)=\phi_0+2\pi t/T$ and $\bar{\theta}(t)=\pi/2$, 
with $T=1$ year, and $\phi_0=0$ specifying the initial orientation with respect to the fixed reference frame. 
In projecting eq.~\eqref{precession} on the unit vector that identifies the binary, we obtain the source's polar 
angle in the detector moving frame as a function of the fixed sky position:
\begin{equation}
\cos\theta_\tn{S}=\frac{1}{2}\cos\bar{\theta}_\tn{S}-\frac{\sqrt{3}}{2}\sin\bar{\theta}_\tn{S}\cos[\bar{\phi}(t)-\bar{\phi}_\tn{S}]\ ,
\end{equation}
and similarly for the azimuth:
\begin{equation}
\bar{\phi}_\tn{S}=\tan^{-1}\left[\frac{\sqrt{3}\cos\bar{\theta}_\tn{S}+\sin\bar{\theta}_\tn{S}\cos[\bar{\phi}(t)-
\bar{\phi}_\tn{S}]}{2\sin\bar{\theta}_\tn{S}\sin[\bar{\phi}(t)-\bar{\phi}_\tn{S}]}\right]\ .\label{phiSD}
\end{equation}
The polarization angle $\psi_\tn{S}$ also changes with time:
\begin{equation}
\psi_\tn{S}(t)=\tan^{-1}\frac{\hat{L}\cdot\hat{z}-(\hat{L}\cdot\hat{N})(\hat{z}\cdot \hat{N})}{\hat{N}\cdot(\hat{L}\times\hat{z})}\ ,
\end{equation}
with $\hat{z}\cdot\hat{N}=\cos\theta_\tn{S}$ and 
\begin{subequations}
\begin{align}
\hat{L}\cdot\hat{z}=&\frac{1}{2}\cos\bar{\theta}_\tn{L}-\frac{\sqrt{3}}{2}\sin\bar{\theta}_\tn{L}\cos[\bar{\phi}(t)-\bar{\phi}_\tn{L}]\ ,\\
\hat{L}\cdot\hat{N}=&\cos\bar{\theta}_\tn{L}\cos\bar{\theta}_\tn{S}+\sin\bar{\theta}_\tn{L}\sin\bar{\theta}_\tn{S}
\cos[\bar{\phi}_\tn{L}-\bar{\phi}_\tn{S}]\ ,\\
\hat{N}\cdot(\hat{L}\times\hat{z})=&\frac{1}{2}\sin\bar{\theta}_\tn{L}\sin\bar{\theta}_\tn{S}
\sin[\bar{\phi}_\tn{L}-\bar{\phi}_\tn{S}]+\nonumber\\
&\frac{\sqrt{3}}{2}\big\{\cos\bar{\theta}_\tn{L}\sin\bar{\theta}_\tn{L}\sin[\bar{\phi}(t)-\bar{\phi}_\tn{S}]\nonumber\\
&-\cos\bar{\theta}_\tn{S}\sin\bar{\theta}_\tn{L}\sin[\bar{\phi}(t)-\bar{\phi}_\tn{L}]\big\}\ .
\end{align}
\end{subequations}
Finally, the Doppler phase shift defined in eq.~\eqref{signal} can be written as 
$\psi_\tn{D}=2\pi f R_\tn{AU}\sin\bar{\theta}_\tn{S}\cos[\bar{\phi}(t)-\bar{\phi}_\tn{S}]$, where $R_\tn{AU}$ 
is the astronomical unit. For DECIGO, we assume that the noise spectral density is given by the following 
expression:
\begin{align}
\frac{S^\tn{D}_h(f)}{\tn{Hz}^{-1}}=7.05\cdot 10^{-48}y&+\frac{4.8\cdot10^{-51}}{y}\left(\frac{f}{\tn{Hz}}\right)^{-4}
\nonumber\\
&+5.33\cdot10^{-52}\left(\frac{f}{\tn{Hz}}\right)^{-4}\ ,
\end{align}
with $y=1+\left(\frac{f}{7.36\tn{Hz}}\right)^2$ \citep{Yagi:2011wg}.

B-DECIGO can be considered as a scaled version of DECIGO, which is build to test the most important 
technological features of the latter. Still, the scientific goals of this reduced mission remain unchanged, with the 
main difference being in a lower detector's sensitivity \citep{Sato:2017dkf,Isoyama:2018rjb}, and therefore a 
smaller number of observations are expected. Although the orbital configuration 
of such a satellite has not been finalized yet, in this paper we assume that B-DECIGO will follow the same 
trajectory\footnote{See \citep{Nair:2018bxj} for a specific example of a geocentric satellite, instead of the one 
considered in this paper that orbits around a Sun centered frame.} of its bigger brother, but with a lower noise spectral density 
given by
\begin{equation}
S^\tn{BD}_h(f)=S_0(1+1.584\cdot 10^{-2}x^{-4}+1.584\cdot 10^{-3}x^{2})\ ,
\end{equation}
where and $x=f/$Hz and $S_0=4.04\times10^{-46}$Hz$^{-1}$ \citep{Isoyama:2018rjb}.

\section{S/N as a function of the source angles}\label{sec:appsnr}

The parameter's space spanned by the GW template \eqref{signal} is rather large, and it is 
not straightforward in analyzing the dependence of the S/N as a function of the 
source's properties. However, since the luminosity distance $d$ acts as a scale factor, for a 
specific choice of the component masses, $\rho$ is only determined by the four angles 
$\vec{\zeta}=(\bar{\theta}_\tn{S},\bar{\phi}_\tn{S},\bar{\theta}_\tn{L},\bar{\phi}_\tn{L})$. 
In this Appendix, we investigate how the S/N varies in terms of $\vec{\zeta}$ for prototype BWDs with 
$m_1=m_1=1M_\odot$ observed by DECIGO and B-DECIGO. We focus on some specific 
choices of the azimuthal and polar angles, studying the dependence on the direction of the 
binary's angular momentum. 

\begin{figure}[th]
\centering
\includegraphics[width=3.9cm]{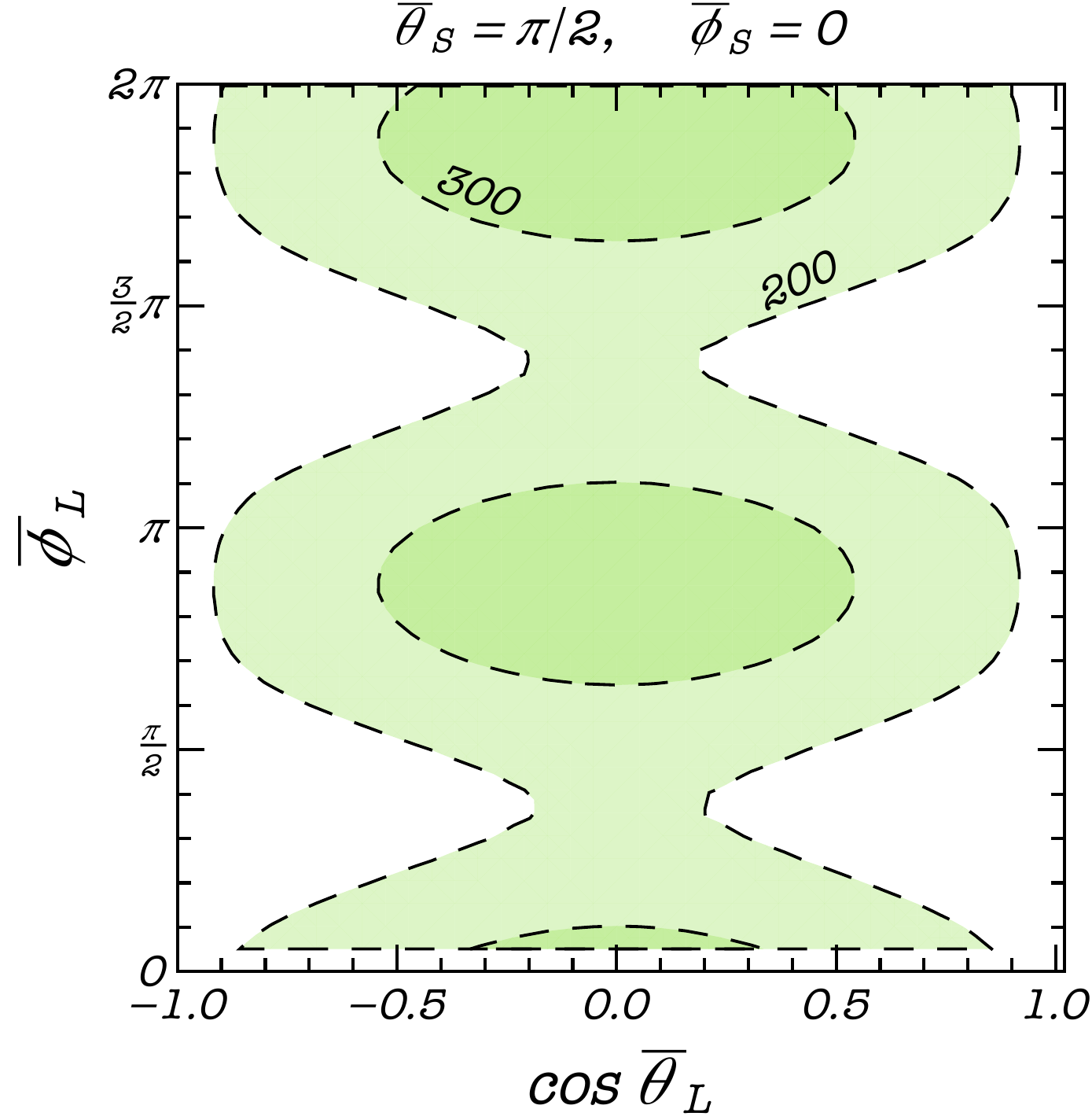}
\includegraphics[width=3.9cm]{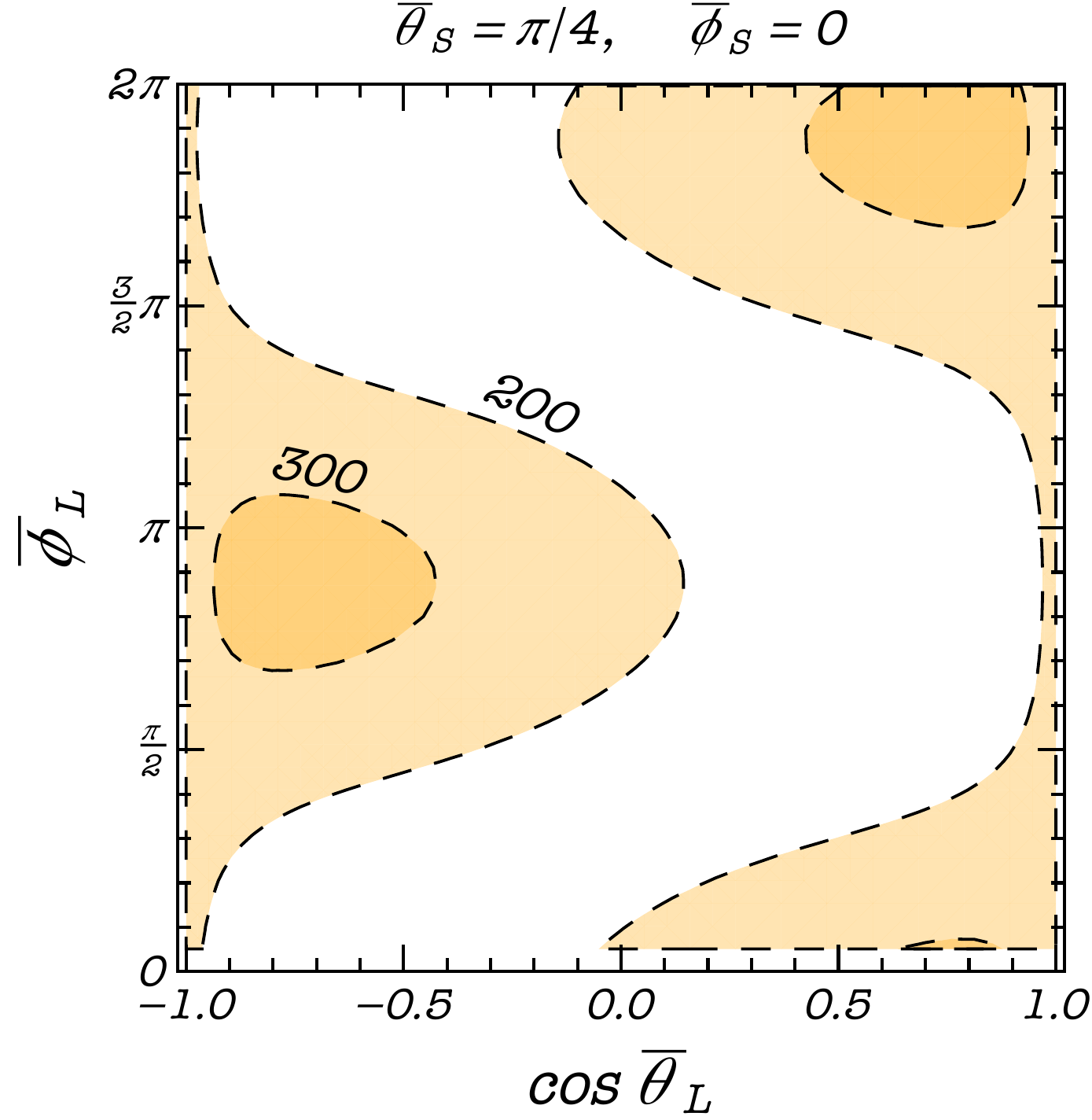}\\
\vspace{0.2cm}
\includegraphics[width=3.9cm]{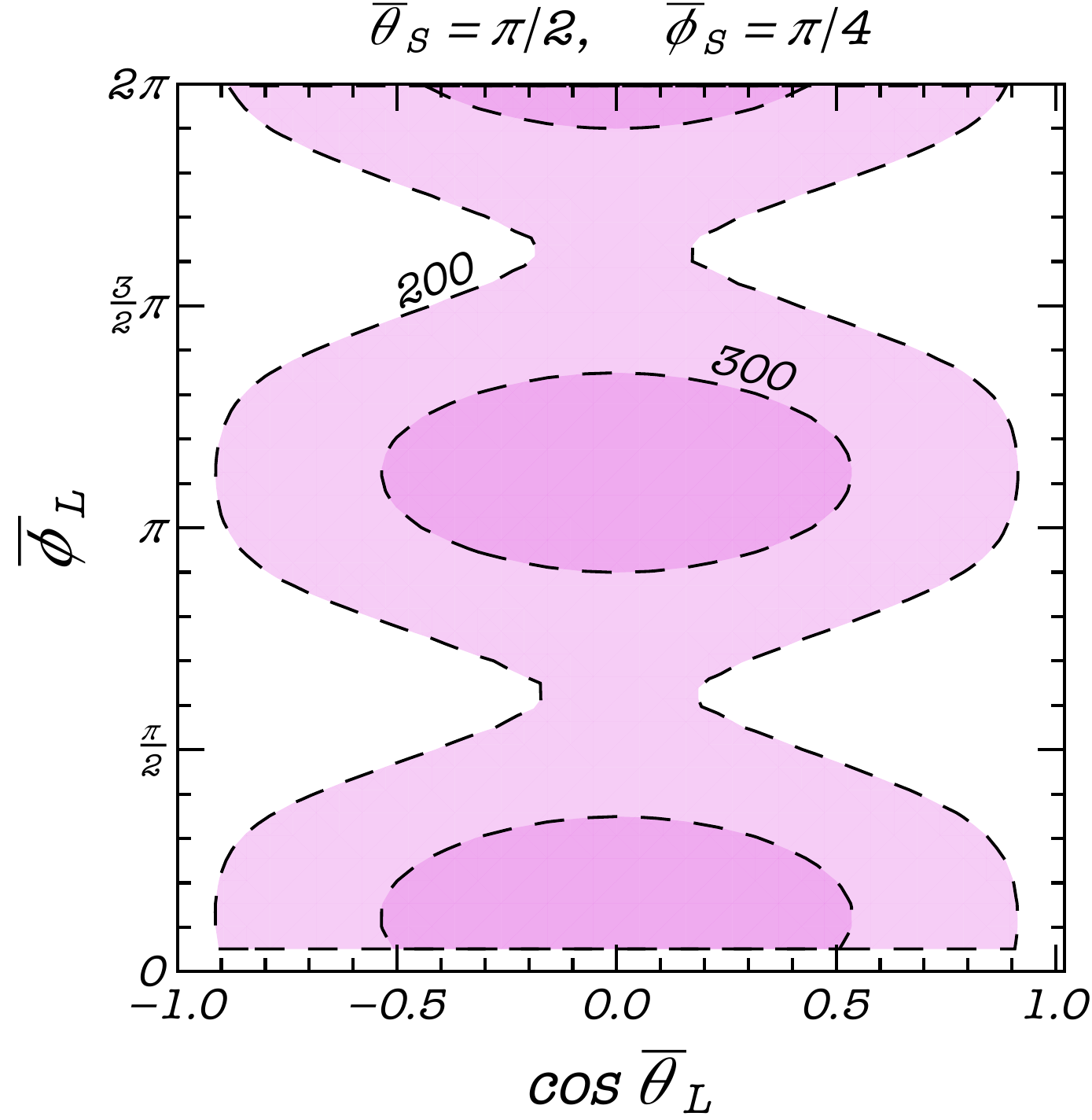}
\includegraphics[width=3.9cm]{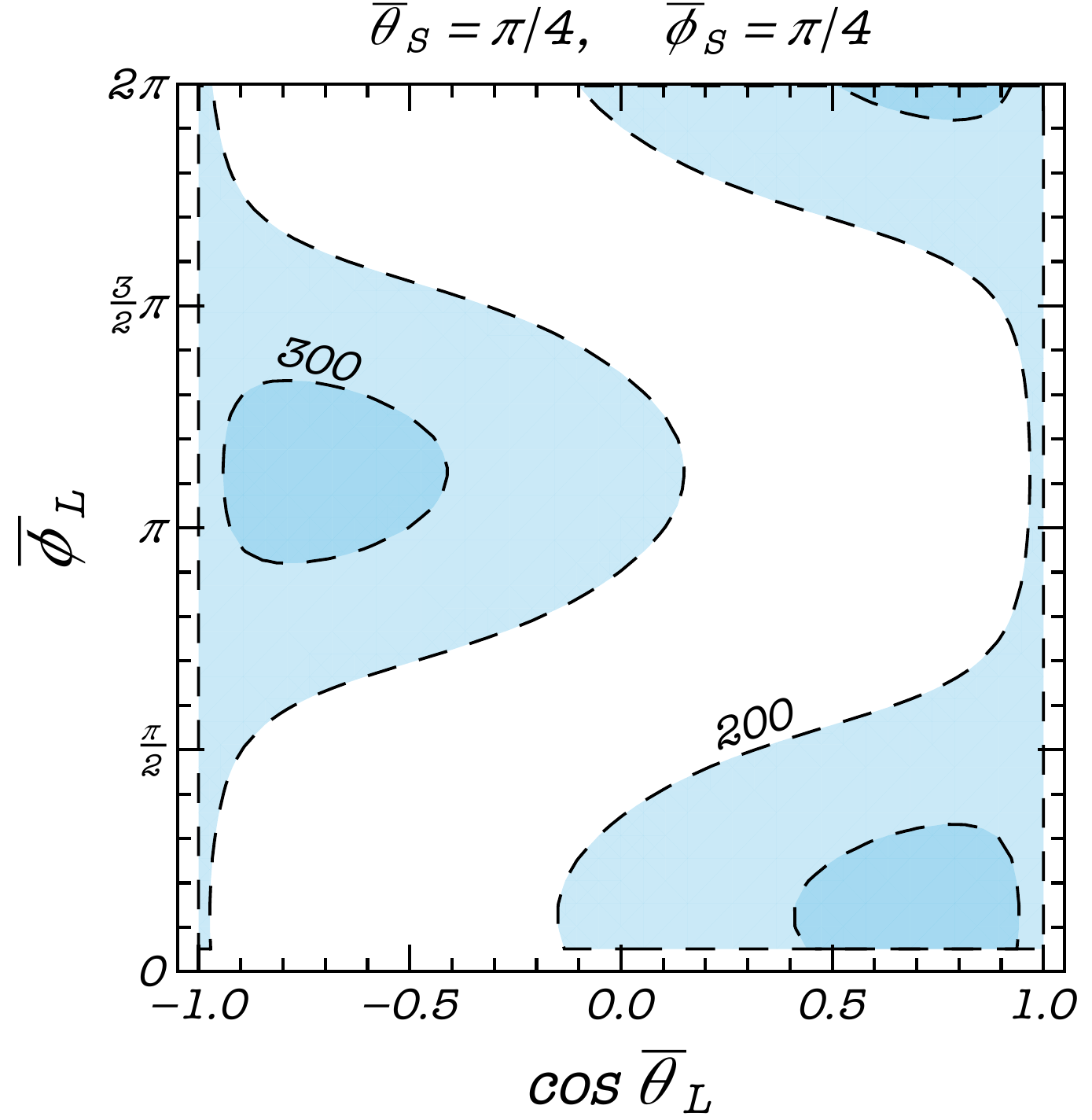}
\caption{Contour regions of S/N for BWD with $m_1=m_2=1M_\odot$ at $d=100$Mpc, 
observed by DECIGO, as a function of the angles that specify the binary angular 
momentum, for a fixed source's orientation in the sky, i.e. $\bar{\theta}_\tn{S} = (\pi/2,\pi/4)$ 
and $\bar{\phi}_\tn{S}=(0,\pi/4)$. Dark (light) contours identify regions where 
$\rho\geq300$ $(\rho\geq200)$.}
\label{fig:snrD} 
\end{figure}

Figure \ref{fig:snrD} shows the regions, for binaries at $d=100$ Mpc detected by DECIGO, 
where $\rho\geq (200,300)$ for $\bar{\phi}_\tn{S} = (0,\pi/4)$ and $\bar{\theta}_\tn{S} = (\pi/2,\pi/4)$. 
As already noted in Sec.~\ref{sec:results} 
for all the configurations, we obtain very high values for the S/N, which are always $\gtrsim 100$ in the 
entire parameter's space and are modulated by the specific direction of the angular momentum 
$\hat{L}$. While $\rho$ seems more sensitive to variations of the source azimuth, different 
choices of the polar angle $\bar{\phi}_\tn{S}$ lead to qualitatively similar results. 

\begin{figure}[th]
\centering
\includegraphics[width=4.2cm]{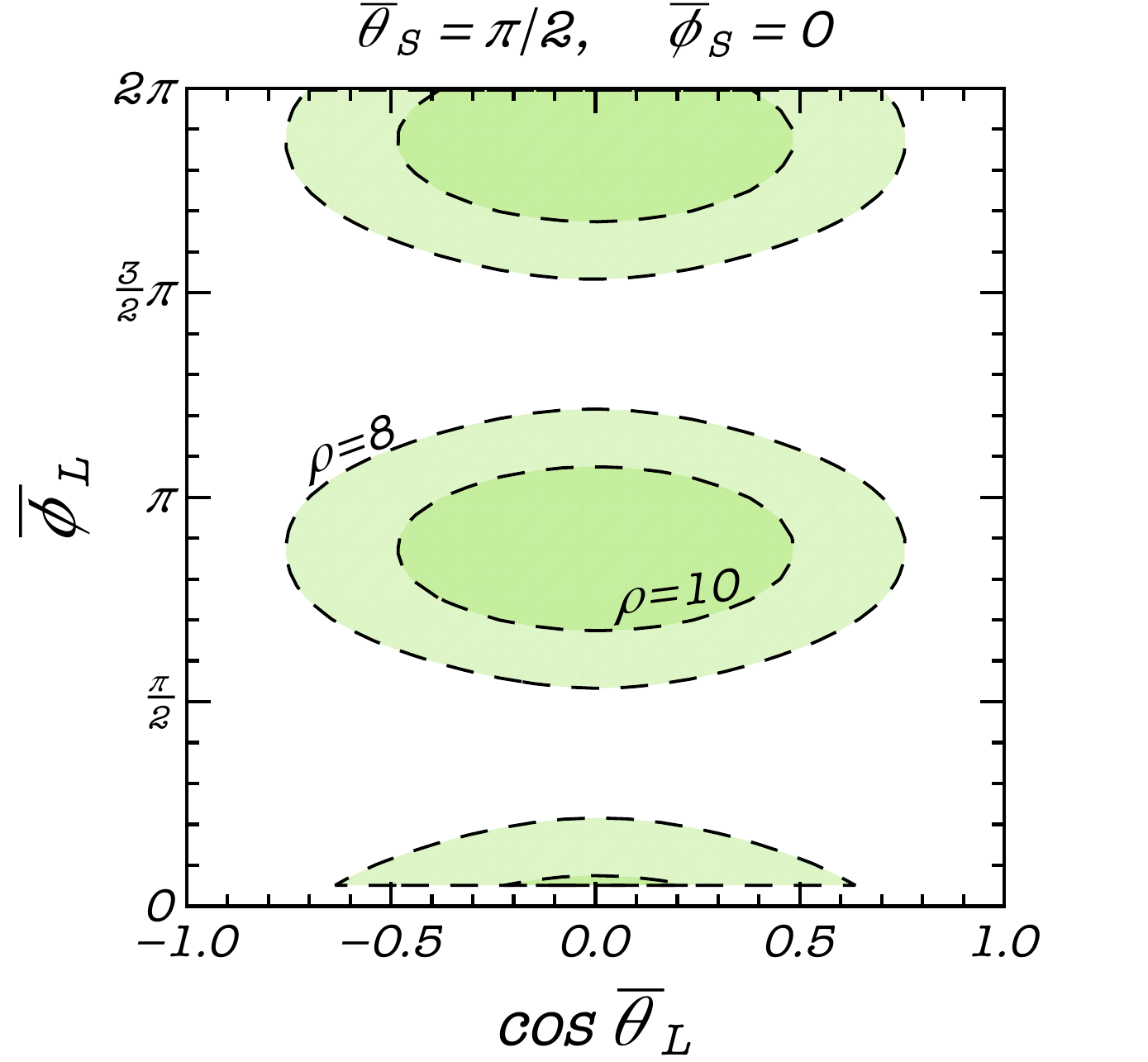}
\includegraphics[width=4.2cm]{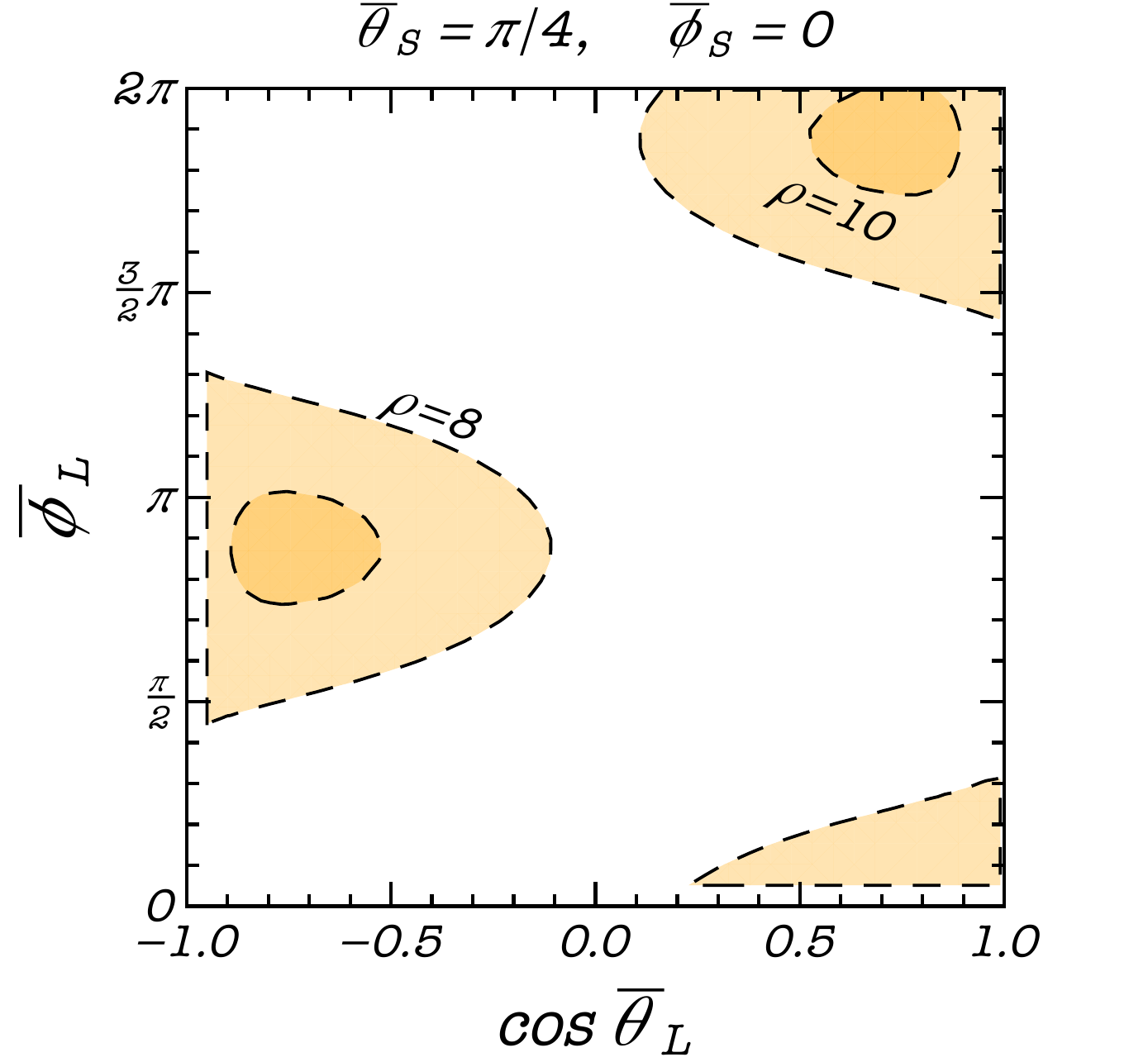}
\includegraphics[width=4.2cm]{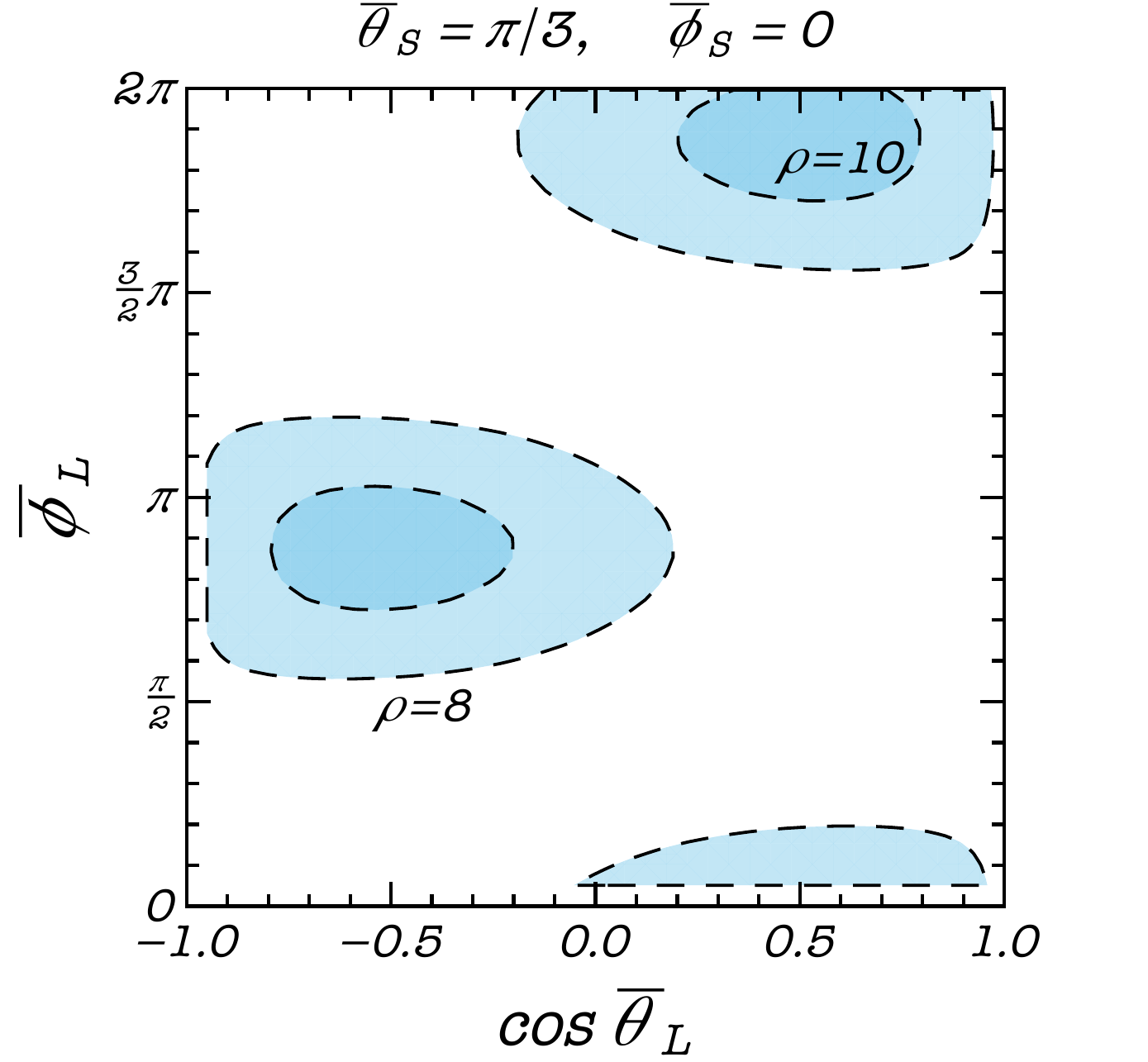}
\caption{Same as Fig.~\ref{fig:snrD}, but for white dwarf binaries detected by B-DECIGO at 
100 Mpc. The shaded regions identify configurations with $\rho\geq 8$ and $\rho\geq 10$.}
\label{fig:snrBD} 
\end{figure}

The three panels of Fig.~\ref{fig:snrBD} show the same analysis for B-DECIGO and BWDs 
at $d=100$Mpc. Colored islands identify the configurations that are observed with 
$\rho\geq10$ and $\rho\geq8$. The latter represents the detection's threshold fixed in 
Sec.~\ref{sec:results}. The shape of the regions resemble the results seen above. However, 
depending on $\bar{\theta}_\tn{L}$ and $\bar{\phi}_\tn{L}$, only specific combinations of 
these two angles yield  binaries observable by the interferometer. These values improve
for sources closer to the detector. For example, at $d=50$ Mpc, all the systems populating 
the plots in Fig.~\ref{fig:snrBD} would be detectable above the threshold.

\end{appendix}

\end{document}